\documentclass[prd,preprint,superscriptaddress,preprintnumbers,eqsecnum,showpacs,nofootinbib,nobibnotes]{revtex4}
\usepackage{amsfonts,bm}
\usepackage{amsfonts,amssymb,amsmath}
\usepackage{graphicx}
\usepackage{paralist}
\usepackage[english]{babel}
\usepackage{slashed}
\usepackage[usenames]{xcolor}
\usepackage{float}
\usepackage{color}

\newcommand{\be}{\begin{equation}}
\newcommand{\bea}{\begin{eqnarray}}
\newcommand{\ee}{\end{equation}}
\newcommand{\eea}{\end{eqnarray}}
\def\n#1{$(#1)$}
\def\s#1{{\scriptscriptstyle #1}}
\def\ie{{\it i.e.}, }
\def\eg{{\it e.g.}, }

\def\1eq#1{Eq.~(\ref{#1})}

\def\2eqs#1#2{Eqs.~(\ref{#1}) and~(\ref{#2})}
\def\3eqs#1#2#3{Eqs.~(\ref{#1}),~(\ref{#2}), and~(\ref{#3})}

\def\Y{Y}

\newcommand{\GT}{{\Gamma}_{\!\!{\s{\mathbf{T}}}}}      

\newcommand{\fatg}{{\rm{I}}\!\Gamma}

\begin{document}

\title{Gluon mass scale through nonlinearities and vertex interplay}

\author{A.~C. Aguilar}
\affiliation{University of Campinas - UNICAMP, 
Institute of Physics ``Gleb Wataghin'',
13083-859 Campinas, SP, Brazil}

\author{M.~N. Ferreira}
\affiliation{University of Campinas - UNICAMP, 
Institute of Physics ``Gleb Wataghin'',
13083-859 Campinas, SP, Brazil}
\affiliation{\mbox{Department of Theoretical Physics and IFIC, 
University of Valencia and CSIC},
E-46100, Valencia, Spain}

\author{C.~T. Figueiredo}
\affiliation{University of Campinas - UNICAMP, 
Institute of Physics ``Gleb Wataghin'',
13083-859 Campinas, SP, Brazil}
\affiliation{\mbox{Department of Theoretical Physics and IFIC, 
University of Valencia and CSIC},
E-46100, Valencia, Spain}

\author{J. Papavassiliou}
\affiliation{\mbox{Department of Theoretical Physics and IFIC, 
University of Valencia and CSIC},
E-46100, Valencia, Spain}

\begin{abstract}

We present a novel analysis of the gluon gap equation,
where its full nonlinear structure is duly taken into account.
In particular, while in previous treatments the linearization of this 
homogeneous integral equation introduced an indeterminacy in the scale of the corresponding mass,
the current approach determines it uniquely, once the value of the gauge coupling at a given
renormalization point is used as input. 
A crucial ingredient for this construction is the ``kinetic term'' 
of the gluon propagator, 
whose form is not obtained from the complicated 
equation governing its evolution, but is rather approximated by
suitable initial {\it Ans\"atze}, 
which are subsequently improved by means of a systematic iterative procedure. 
The multiplicative renormalization of the central  
equation is carried out following an approximate method, which is 
extensively employed in the studies of the standard quark gap equation. 
This approach amounts to the effective substitution of the 
vertex renormalization constants by 
kinematically simplified form factors of the three- and four-gluon vertices.
The resulting numerical interplay, exemplified by the infrared suppression
of the three-gluon vertex and the mild enhancement
of the four-gluon vertex, is instrumental for obtaining
positive-definite and  monotonically decreasing running gluon masses. 
The resulting gluon propagators, put together from the
gluon masses and kinetic terms obtained with this method,
match rather accurately the data obtained from large-volume lattice simulations.

\end{abstract}


\maketitle

\section{\label{sec:intro} Introduction}
The nonperturbative aspects of the gluon propagator, both in pure Yang-Mills theory and in QCD, are believed to be relevant for our understanding of a wide range of important physical  phenomena, such as the dynamical generation of a mass gap, confinement, chiral symmetry breaking, and the formation of bound states such as mesons, baryons, glueballs, hybrids, and exotics~\cite{Roberts:1994dr,Alkofer:2000wg,Maris:2003vk,Greensite:2003bk,Fischer:2006ub,Pawlowski:2005xe,Binosi:2009qm,Boucaud:2011ug,Vandersickel:2012tz,Maas:2011se,Cloet:2013jya,Meyer:2015eta,Kondo:2014sta,Aguilar:2015bud,Eichmann:2016yit,Cyrol:2017ewj,Huber:2018ned}. 
A particularly interesting feature of the gluon propagator, which manifests itself both in the Landau gauge and away from it, is the saturation of its scalar form factor, $\Delta(q^2)$, in the deep infrared, \ie  $\Delta(0) = c >0$. This special behavior, which is intimately connected with the emergence of a gluon mass scale~\cite{Cornwall:1981zr,Bernard:1981pg,Bernard:1982my,Donoghue:1983fy,Wilson:1994fk,Philipsen:2001ip,Aguilar:2002tc},
has been firmly established in a variety of SU(2)~\cite{Cucchieri:2007md,Cucchieri:2007rg,Cucchieri:2009zt,Cucchieri:2009kk} and SU(3)~\cite{Bogolubsky:2007ud,Bowman:2007du,Bogolubsky:2009dc,Oliveira:2009eh,Ayala:2012pb,Bicudo:2015rma} large-volume lattice simulations, and has been extensively studied in the continuum within diverse theoretical frameworks~\mbox{\cite{Aguilar:2004sw,Aguilar:2006gr,Braun:2007bx,Epple:2007ut,Aguilar:2008xm,Fischer:2008uz,Boucaud:2008ky,Dudal:2008sp,RodriguezQuintero:2010wy,Campagnari:2010wc,Tissier:2010ts,Pennington:2011xs,Serreau:2012cg,Fister:2013bh,Binosi:2014aea,Cyrol:2014kca,Meyers:2014iwa,Siringo:2015wtx,Aguilar:2016ock,Glazek:2017rwe,Cyrol:2018xeq}.}

One of the approaches put forth in order to explain the infrared saturation of $\Delta(q)$
relies on the implementation  
of   the  Schwinger   mechanism~\cite{Schwinger:1962tn,Schwinger:1962tp} at the
level of the Schwinger-Dyson equation (SDE) that controls the momentum evolution of $\Delta(q)$. 
This SDE, in turn, has been formulated within  the framework developed through the
merging of the pinch-technique
(PT)~\cite{Cornwall:1981zr,Cornwall:1989gv,Pilaftsis:1996fh,Binosi:2009qm}
with  the  background-field method
(BFM)~\cite{DeWitt:1967ub,Honerkamp:1972fd,Kallosh:1974yh,KlubergStern:1974xv,Abbott:1980hw}, known as
the ``PT-BFM   scheme''~\cite{Aguilar:2006gr,Binosi:2007pi}.
$\Delta(q)$  is subsequently written as the sum of two distinct components,
the kinetic term, $J(q)$, and the mass term, $m^2(q)$, according to \1eq{eq:gluon_m_J}.
This splitting enforces a special realization of the Slavnov-Taylor identity (STI) 
satisfied by the fully dressed three-gluon vertex entering in the gluon SDE [see \1eq{stiv}], 
leading finally to the separation of this dynamical equation into a a system of two coupled integral equations,
one for each component~\cite{Binosi:2012sj}.

Even though the derivation of the aforementioned system is theoretically well-defined,
its complete treatment is still pending, mainly due to the technical complexities associated with the equation governing $J(q)$.
Instead, one considers only the homogeneous integral equation for $m^2(q)$, whose form is given by
(\mbox{$\alpha_s:= g^2/4\pi$}, and $g$ denotes the gauge coupling)
\be
m^{2}(q) = \int_{k} m^2(k) \Delta(k)\Delta(k+q){\cal K} (k,q,\alpha_s),
\label{meqgen}
\ee
and solves it in isolation~\cite{Binosi:2012sj,Aguilar:2014tka}.
In particular, the propagators $\Delta$ appearing in \1eq{meqgen}
are {\it not} decomposed according to  
\1eq{eq:gluon_m_J} but are rather treated as {\it external} quantities, whose form is taken from the data of
large-volume lattice simulations.
This practical simplification, however, distorts the true nature of the original equation,   
converting it to a {\it linear} integral equation. As a consequence,
one has to deal with an eigenvalue problem, which has a solution for a {\it unique}  value of $\alpha_s$,
rather than a continuous interval of values.
Moreover, due to its linearity, the equation 
admits a family of solutions parametrized by a real constant $c_0>0$, since, 
if $m^{2}(q)$ is a solution, so is $c_0\times m^{2}(q)$ (for that unique $\alpha_s$). 
This fact, in turn, introduces an ambiguity in the physics, 
because the final scale must be introduced ``by hand'',
with no clear connection to the fundamental parameters 
of the theory. 

Evidently, it would be far preferable to work with a dynamical equation that allows one 
to determine
how the emergent scale responds to changes in the value of $\alpha_s$ at a given scale $\mu$,
furnishing the ``correct'' mass (\ie the one set by the lattice) 
once a special value for $\alpha_s$ has been chosen.
In that sense, one is seeking to replicate the circumstances that occur in the 
context of the quark gap equation, 
where $\alpha_s$ may be varied, within a reasonable range, giving rise
to a continuous set of quark masses; and once the value of the quark mass has been
fixed with a given accuracy (say, from the lattice), a firm restriction 
on the allowed values for $\alpha_s$ may be obtained.
As we will show in the present work, this is indeed what happens 
after the nonlinear nature of the original gluon mass equation has been restored.

In practice, the restoration of the nonlinearity of \1eq{meqgen} 
is accomplished by implementing in it the substitution \1eq{eq:gluon_m_J}, using a set of
physically motivated {\it Ans\"atze} for $J(q)$.
Specifically, even in the absence of a full treatment of the corresponding dynamical equation, the preeminent qualitative features
of $J(q)$ are relatively well-known,
due to its profound relation with the three-gluon vertex~\cite{Aguilar:2013vaa,Aguilar:2019jsj}.
In particular, as the Euclidean momentum $q^2$ decreases,  
$J(q)$ departs  gradually from its tree-level value,   
reverses its sign (``zero-crossing''), and finally diverges 
logarithmically at the origin.
In fact, as has been argued in the works cited above, these special properties of
$J(q)$ are inextricably connected with the infrared suppression displayed by the main form factors of the
three-gluon vertex~\cite{Cucchieri:2006tf,Cucchieri:2008qm,Huber:2012kd,Pelaez:2013cpa,Blum:2014gna,Eichmann:2014xya,Vujinovic:2014fza,Athenodorou:2016oyh,Duarte:2016ieu,Boucaud:2017obn}. 
When a $J(q)$ that encodes the above features is used as an initial ``seed'',
and a value for $\alpha_s$ is chosen, 
the gluon mass equation yields a unique  $m^{2}(q)$. The procedure is further refined
by modifying the shape of $J(q)$ and by adjusting\footnote{As we will see in Sec.~\ref{sec:numan}, a precision of about 1\% is required.} the value of $\alpha_s$, 
such that the resulting propagator,
obtained by combining $J(q)$ and $m^{2}(q)$ according to \1eq{eq:gluon_m_J},
matches the lattice result as accurately as possible.

There is an additional issue that appears when dealing with the gluon mass equation~\eqref{meqgen}, 
related with the relative size of the one-loop and
two-loop dressed contributions, which enter in the kernel ${\cal K} (k,q,\alpha_s)$ 
with a relative minus sign.
Specifically, a positive-definite and monotonically decreasing
solution for $m^{2}(q)$ requires a delicate balance between these two terms,
which depends, among other things, on the
way that the multiplicative renormalization of the equation is enforced. 
In the case of the  quark gap equation, this problem has 
been dealt with by means of an approximate method, which amounts to the
substitution of the renormalization constants
by appropriately chosen momentum-dependent functions~\cite{Fischer:2003rp,Aguilar:2010cn,Aguilar:2018epe,Huber:2018ned}. In this work we resort to the same expedient, appropriately adapted to the
specific vertices appearing in the problem. 
It turns out that its implementation
introduces a subtle interplay between the
three- and four-gluon vertices, which is instrumental for the overall stability of the 
resulting integral equation.

The article is organized as follows.
In Sec.~\ref{sec:overview} we review the most salient features of the
integral equation that governs the existence and momentum evolution of $m^2(q^2)$.  
Then, in Sec.~\ref{sec:multren} we present the procedure adopted for the
effective implementation of the multiplicative renormalization, drawing an analogy with the
more familiar case of the quark gap equation, and elaborating on the main 
underlying assumptions.
In Sec.~\ref{sec:inp} we discuss in detail the origin and properties of the various
ingredients entering in the kernel of the gluon mass equation.
In Sec.~\ref{sec:numan} we present a thorough numerical study
of the resulting integral equation, and obtain solutions for $m^2(q^2)$ that reproduce quite accurately
the saturation scale of the gluon propagator observed in lattice simulations, for values
of $\alpha_s$ that are in accordance with the theoretical expectations. 
Finally, in Sec.~\ref{sec:conc} we discuss our results and comment on further possible directions.

\section{\label{sec:overview} Gluon mass equation: a brief overview}

In this section we briefly review the structure of the gluon mass equation, and discuss
in some detail its multiplicative renormalization. 

\subsection{Basic concepts and ingredients}

Throughout this article we work in the {\it Landau gauge}, where the 
gluon propagator $\Delta^{ab}_{\mu\nu}(q)=-i\delta^{ab}\Delta_{\mu\nu}(q)$ is completely transverse, given by 
\begin{align}
\Delta_{\mu\nu}(q) = \Delta(q) {\rm P}_{\mu\nu}(q)\,, \qquad {\rm P}_{\mu\nu}(q) = g_{\mu\nu} - \frac{q_\mu q_\nu}{q^2}\,.
\end{align}
The special property of infrared saturation displayed by $\Delta(q)$ prompts its splitting
into two separate components, according to (Euclidean space)~\cite{Binosi:2012sj} 
\be
\label{eq:gluon_m_J}
\Delta^{-1}(q) = q^2J(q) + m^2(q)\,,
\ee
where $J(q)$ corresponds to the so-called ``kinetic term''
(at tree-level, $J(q) =1$), 
while $m^2(q)$ to a momentum-dependent
gluon mass scale, with the property $m^2(0)=\Delta^{-1}(0)$.
For large values of $q^2$, the component $J(q)$ captures 
standard perturbative corrections to the gluon propagator, while in the infrared it 
exhibits several exceptional characteristics~\cite{Aguilar:2013vaa,Aguilar:2019jsj}.

\begin{figure}[t]
\begin{minipage}[c]{0.45\linewidth}
\centering 
\hspace{-1.5cm}
\includegraphics[scale=0.4]{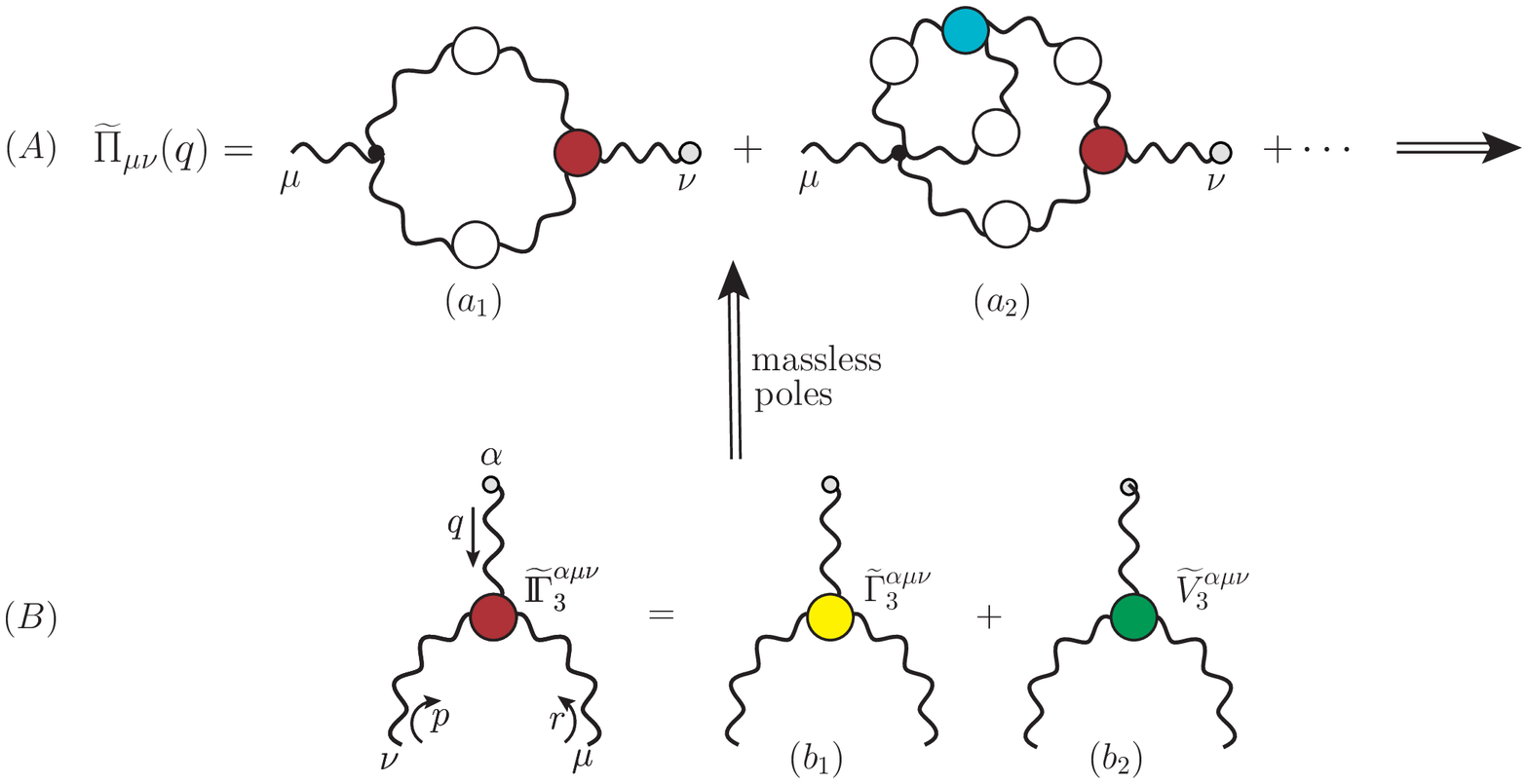}
\end{minipage}
\hspace{-0.3cm}
\begin{minipage}[c]{0.4\linewidth}
\includegraphics[scale=0.37]{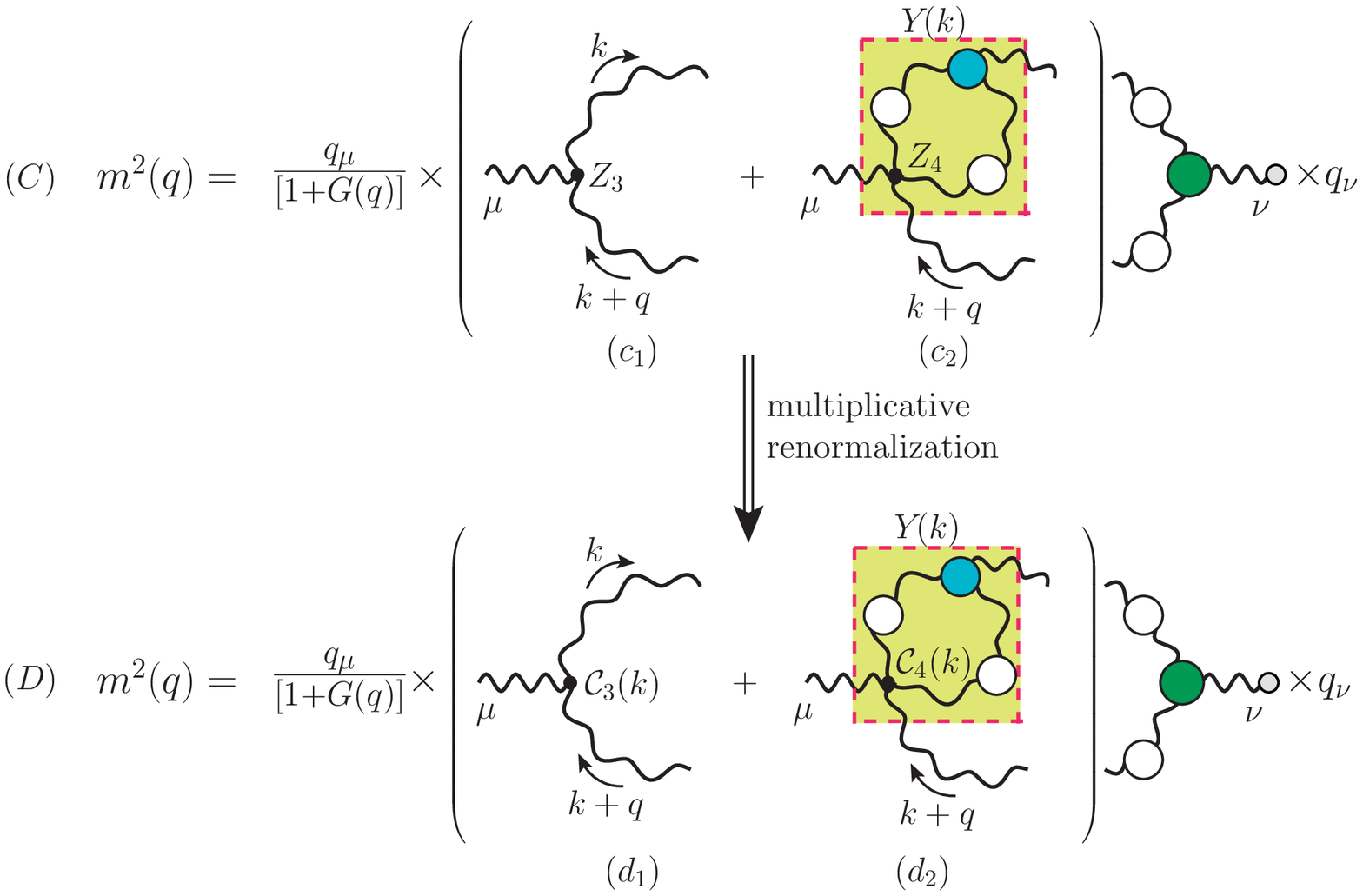}
\end{minipage}
\caption{The schematic representation of the steps involved in the derivation of the dynamical gluon mass equation in the PT-BFM framework. White (colored)
 circles denote fully dressed propagators (vertices).}
\label{fig:masseq}
\end{figure}

The full dynamical evolution of the functions 
$J(q)$ and $m^2(q)$ is determined by two separate, but coupled, integral equations,
whose derivation may be carried out within the PT-BFM framework~\cite{Aguilar:2006gr,Binosi:2007pi,Binosi:2008qk}.
To that end, the most advantageous starting point 
is the mixed propagator connecting a quantum ($Q$) and a background ($B$) gluon, 
to be denoted by  $\widetilde{\Delta}(q^2)$; the diagrammatic representation
of its self-energy, ${\widetilde\Pi}_{\mu\nu}(q)$, is given in panel $(A)$ of Fig.~\ref{fig:masseq}.
When contracted by the momentum $q^{\nu}$, 
the fully dressed vertices appearing in ${\widetilde\Pi}_{\mu\nu}(q)$
satisfy Abelian STIs; for instance, the $BQQ$ vertex  appearing in the panel $(B)$ of Fig.~\ref{fig:masseq} satisfies  
\be 
\label{WIBQ2}
q_\alpha {\widetilde\fatg}_{\!3}^{\alpha\mu\nu}(q,r,p) =
\Delta^{-1}(p) {\rm P}^{\mu\nu}(p) -  \Delta^{-1}(r) {\rm P}^{\mu\nu}(r)\,. 
\ee
This property, in turn, makes the realization of the transversality condition \mbox{$q^{\nu}{\widetilde\Pi}_{\mu\nu}(q)=0$}
considerably more transparent. Note also that the conventional ($QQ$) gluon propagator, $\Delta(q^2)$, 
is connected to $\widetilde{\Delta}(q^2)$ by  
the exact relation~\cite{Grassi:1999tp, Binosi:2002ez},
\be
\Delta(q) = [1 + G(q)] \widetilde{\Delta}(q)\,,
\label{BQIs}
\ee
where $1+G(q)$ is the $g_{\mu\nu}$ co-factor of a special two-point correlation function
(see~\cite{Aguilar:2009nf} and references therein), intrinsic to the Batalin-Vilkovisky formalism. Thus, the SDE of interest reads
\be
[q^2J(q) + m^2(q)]{\rm P}_{\mu\nu}(q) = 
\frac{q^2 {\rm P}_{\mu\nu}(q) + i\,{\widetilde\Pi}_{\mu\nu}(q)}{1+G(q^2)}\,.
\label{sde}
\ee
As has been explained in detail in a series of works (see, \eg ~\cite{Aguilar:2011xe,Aguilar:2016vin})  
the emergence of an infrared finite solution for $\Delta$ proceeds through the
activation of the {\it Schwinger mechanism} 
by {\it longitudinally coupled massless poles} contained in the vertex ${\widetilde\fatg}_{\!3}$~\cite{Schwinger:1962tn,Schwinger:1962tp,Jackiw:1973tr,Smit:1974je,Eichten:1974et,Poggio:1974qs}.
Specifically, one separates the three-gluon vertex ${\widetilde\fatg}_{\!3}$ into two distinct parts, 
\be
\label{eq:Gnp}
{\widetilde\fatg}_{\!3}^{\alpha\mu\nu}(q,r,p) ={\widetilde\Gamma}_{\!3}^{\alpha\mu\nu}(q,r,p) + {\widetilde V}_{\!3}^{\alpha\mu\nu}(q,r,p)\,,
\ee
where ${\widetilde\Gamma}_{\!3}$
is pole-free\footnote{Note, however, that it contains logarithmic infrared divergences~\cite{Aguilar:2013vaa,Aguilar:2019jsj}.}, 
while ${\widetilde V}_{\!3}^{\alpha\mu\nu}(q,r,p)$ denotes the part containing the massless bound-state excitations~\cite{Aguilar:2011xe} [see panel $(B)$ in Fig.~\ref{fig:masseq}].
These two components of the full vertex contribute to the realization of \1eq{WIBQ2} in a very particular way:
when the terms $\Delta^{-1}$ on the r.h.s. of~\1eq{WIBQ2} are written in the form of \1eq{eq:gluon_m_J}, then
the divergence of 
$\widetilde{\Gamma}_{\!3}$ on the l.h.s. accounts for the appearance of the $J$ terms,
while ${\widetilde V}_{\!3}$ for the masses, \ie
\bea
q_{\alpha} \widetilde{\Gamma}_{\!3}^{\alpha\mu\nu}(q,r,p) &=&  p^2 J(p) {\rm P}^{\mu\nu}(p) - r^2 J(r) {\rm P}^{\mu\nu}(r) \,,
\nonumber\\
q_\alpha {\widetilde V}_{\!3}^{\alpha\mu\nu}(q,r,p) &=&  m^2(p) {\rm P}^{\mu\nu}(p) - m^2(r) {\rm P}^{\mu\nu}(r) \,.
\label{stiv}
\eea
From this point on, the derivation of the equations for $m^2(q)$ and $J(q)$  
proceeds by associating the pole related parts of each diagram on the r.h.s. of \1eq{sde}
with the $m^2(q)$ term on the l.h.s., assigning the remaining
pieces to the equation for $J(q)$ [see panel $(C)$ in Fig.~\ref{fig:masseq}]. Focusing on the former case, 
after certain algebraic manipulations~\cite{Binosi:2012sj}, the integral equation
that controls the evolution of $m^2(q)$ is given by 
\be
m^2(q) = \frac{g^2 C_A}{1+G(q)}\frac{1}{q^2}\int_k m^2(k)\Delta(k)\Delta(k+q){\cal K}_{m}(q,k)\,,
\label{fullmass} 
\ee
where $C_A$ represents the Casimir eigenvalue of the adjoint representation [$N$ for SU($N$)],
the kernel ${\cal K}_{m}(q,k)$ is given by
\be
{\cal K}_{m}(q,k) = \left\{ {\cal K}^{+}(q,k)[(k+q)^2 - k^2] g^{\alpha\beta}
+ {\cal K}^{-}(q,k)(q^2 g^{\alpha\beta} - 2 q^{\alpha} q^{\beta})\right\}  {\rm P}_{\alpha}^{\rho}(k){\rm P}_{\beta\rho}(k+q) \,,
\ee
with 
\bea
{\cal K}^{+}(q,k) &=& [\Y(k+q) + \Y(k)] - 1 \,,
\nonumber\\
{\cal K}^{-}(q,k) &=& [\Y(k+q) - \Y(k)] \,,
\label{kernels}
\eea
and we have defined
\be
\int_k := \int \frac{\mathrm{d}^4 k}{(2\pi)^4} \,.
\ee
Finally, the function $\Y(k)$ originates from
the subgraph shown in the two-loop diagram $(c_2)$ of Fig.~\ref{fig:masseq}; its closed expression is given in \1eq{Ydef}.

\subsection{Renormalization of the gluon mass equation}

At the formal level, the renormalization of \1eq{fullmass} is carried out multiplicatively,
through the introduction of the appropriate wave-function, vertex, and gauge coupling
renormalization constants. Specifically, 
the fully dressed renormalized quantities (carrying the index ``${R}$'')
are related to the bare ones through~\cite{Aguilar:2014tka}
\bea
\Delta_{\s R}(q) &=&  Z^{-1}_{\s A} \Delta(q)\,, 
\nonumber\\
1+G_{\s R}(q) &=& Z_{\s G}[1+G(q)]\,,
\nonumber\\
\Gamma^{\mu\alpha\beta}_{\!3\s R}(q,r,p) &=& Z_3 \,\Gamma_{\!3}^{\mu\alpha\beta}(q,r,p) \,,
\nonumber\\
g_{\s R} &=& Z^{-1}_g g \,,
\label{renconst2}
\eea
where all renormalization constants $Z_i$ depend both on the ultraviolet cutoff and 
the renormalization point $\mu$.
In what follows we  employ the momentum subtraction (MOM) scheme~\cite{Weinberg:1951ss,Celmaster:1979km}; 
propagators assume their tree-level values at the subtraction point  $\mu$, while 
an analogous condition is imposed on the vertices at special momentum configurations,
such as the ``symmetric'' point. 

Then, the renormalization of \1eq{fullmass} 
is carried out by replacing the bare quantities appearing in them by their renormalized counterparts,
according to \1eq{renconst2}. Specifically,  
suppressing all momentum arguments and indices,
omitting the integral signs  $\int_k$ and $\int_\ell$, and setting $Y \sim g^2 \Delta^2 \Gamma_3$ [see \1eq{Ydef}],
we have
\bea
g^2  \Delta^2 \,[1+G]^{-1} &=& Z_3\, g^2_{\s R}\, \Delta^2_{\s R}\, [1+G_{\s R}]^{-1} \,,
\nonumber\\
g^4 \Delta^{4}  \,\Gamma_3 \,[1+G]^{-1} &=& Z_4\, g^4_{\s R}\, \Delta^4_{\s R} \,\Gamma_{\!3\s R} \, [1+G_{\s R}]^{-1}\,.
\label{renstrings}
\eea
In deriving the above results we have 
used the crucial constraints that 
the fundamental STIs of the theory impose on the
renormalization constants, namely 
\be
Z_g = Z_3  Z_{\s A}^{-3/2} = Z^{-1}_{\s G} {Z}^{-1/2}_{\s A} =Z_4^{1/2} Z_{\s A}^{-1},
\label{STIrel}
\ee
where the last relation involves the four-gluon vertex renormalization
constant, $Z_4$, defined as (suppressing color indices)
\be
\Gamma^{\mu\nu\rho\sigma}_{\!4 \s R}(q,r,p,t) = Z_4 \,\Gamma_{\!4}^{\mu\nu\rho\sigma}(q,r,p,t)\,.
\label{4gren}
\ee
Note that
\n{i} the fully dressed vertex $\Gamma_{\mu\nu\rho\sigma}(q,r,p,t)$ does {\it not} appear
in \1eq{fullmass}; only its tree-level version, $\Gamma_{\mu\nu\rho\sigma}^{(0)}$, 
appears in graph $(c_2)$ of Fig.~\ref{fig:masseq}, and \n{ii} the second relation of \1eq{STIrel}
originates from the fact that, due to the special properties of the PT-BFM framework, 
the combination~\cite{Aguilar:2009nf,Binosi:2014aea,Binosi:2016nme}
\be
{\cal R}_{\s G} = g \Delta^{1/2}(q) [1+G(q)]^{-1} = g_{\s R} \Delta^{1/2}_{\s R}(q) [1+G_{\s R}(q)]^{-1},
\label{theRG}
\ee
is renormalization-group invariant (RGI) [independent of the choice of the
renormalization (subtraction) scale $\mu$, and the ultraviolet cutoff $\Lambda$]. 

Armed with the above relations,
it is relatively straightforward to establish that the net effect of renormalization amounts to the
replacement of bare by renormalized quantities on both sides of \1eq{fullmass},
together with the  modification of the kernels of \1eq{kernels} into 
\bea
{\cal K}^{+}_{\s R}(q,k) &=& Z_4 [\Y_{\s R}(k+q) + \Y_{\s R}(k)] - Z_3 \,,
\nonumber\\
{\cal K}^{-}_{\s R}(q,k) &=& Z_4 [\Y_{\s R}(k+q) - \Y_{\s R}(k)]\,,
\label{renkernel}
\eea
as illustrated in panel $(C)$ of Fig.~\ref{fig:masseq}.

\section{\label{sec:multren} Effective treatment of multiplicative renormalization}

The rigorous implementation of multiplicative renormalization
at the level of SDEs is known to be an exceptionally complicated issue,
which, at the practical level, is resolved by means of certain approximate approaches (see, \eg~\cite{Bloch:2001wz,Bloch:2002eq}).  
In this section we present an effective treatment of this problem,
whose origin may be traced back to analogous
approaches implemented in the studies of the quark
gap equation~\cite{Fischer:2003rp,Aguilar:2010cn,Aguilar:2018epe} and the SDEs 
of vertices~\cite{Huber:2018ned}. The upshot of
this analysis is that the constants $Z_3$ and $Z_4$ in \1eq{renkernel} will be replaced
by appropriate form factors of the three- and four-gluon vertices, respectively.

\subsection{\label{sec:csb}The quark gap equation paradigm}

\begin{figure}[t]
\begin{minipage}[b]{0.45\linewidth}
\centering 
\hspace{-1.5cm}
\includegraphics[scale=0.4]{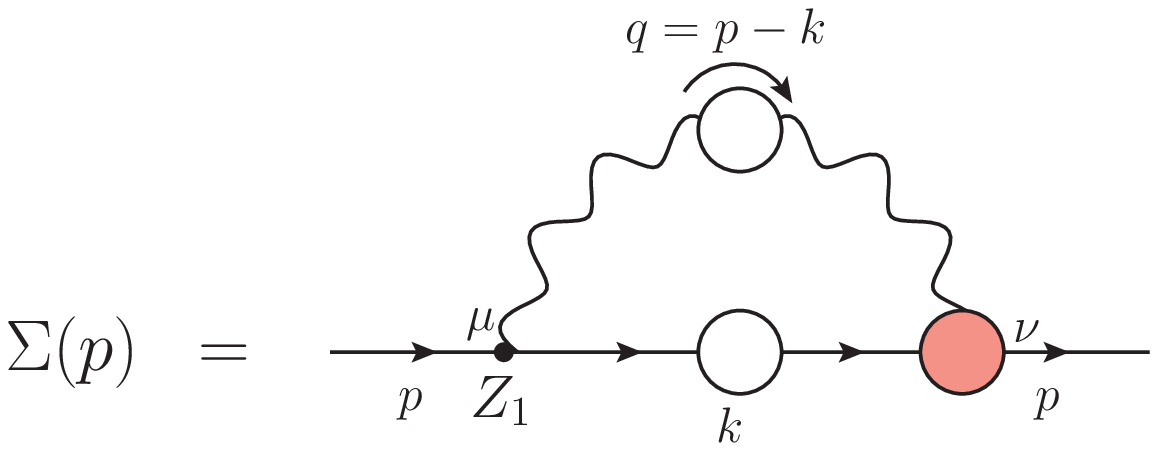} \\
\vspace{1.0cm}
\includegraphics[scale=0.4]{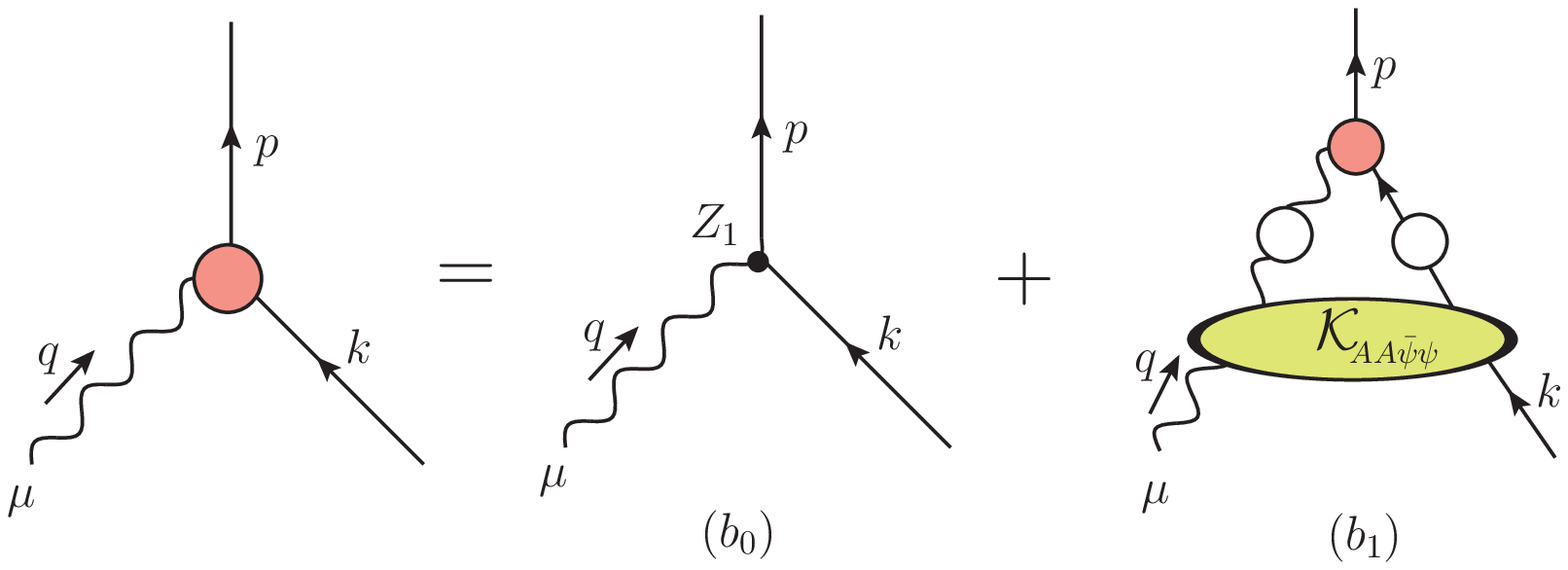}
\end{minipage}
\begin{minipage}[b]{0.45\linewidth}
\includegraphics[scale=0.20]{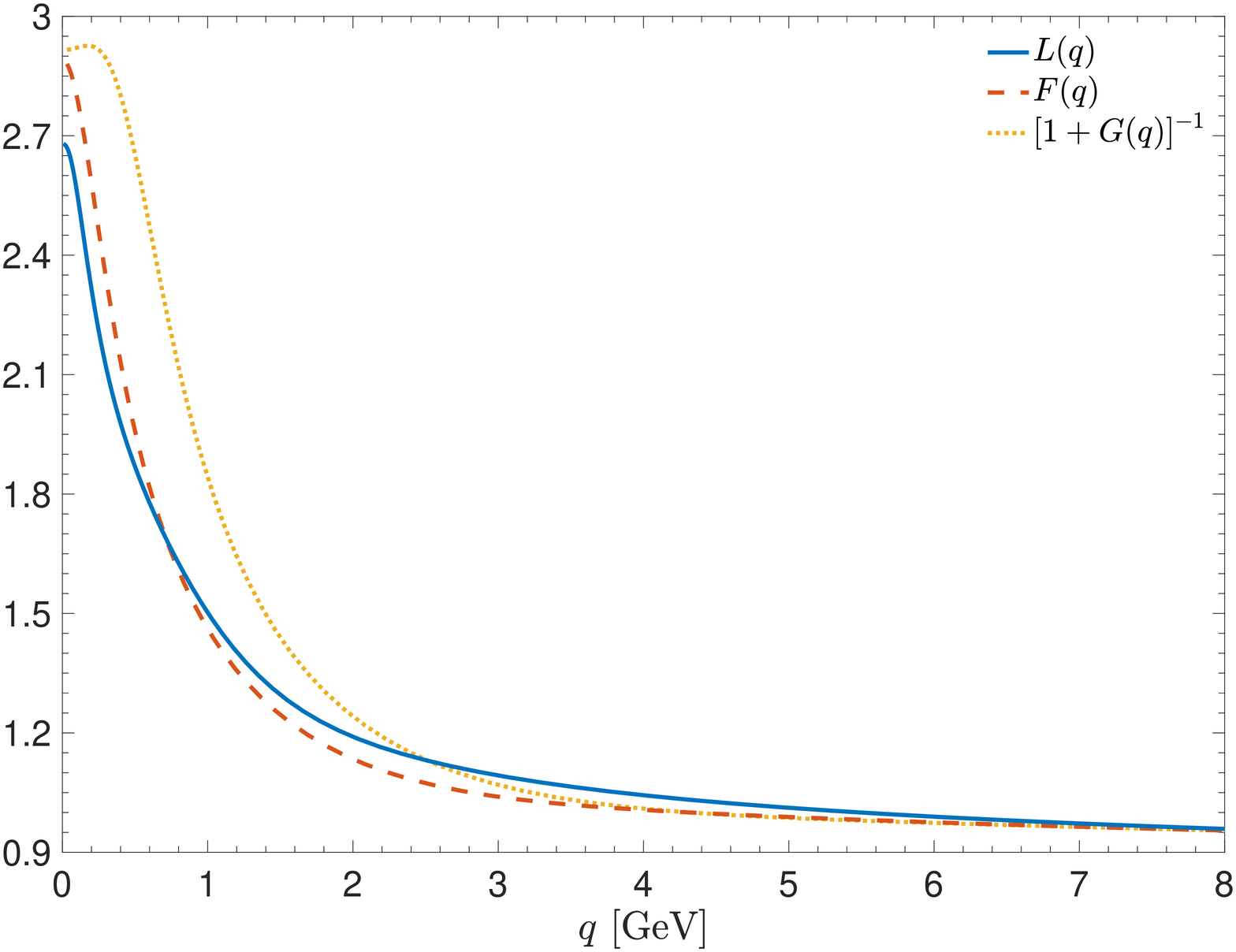}
\end{minipage}
\caption{Left panel: The quark gap equation (top) and the SDE for the quark-gluon vertex $\Gamma^{\mu}$ (bottom),
  expressed with the quark providing the ``reference'' leg~\cite{Binosi:2016rxz}. Right panel:   The three versions of ${\mathcal C}_1(q)$ listed in Eq.~\eqref{threeC}.}
\label{fig:quark}
\end{figure}

It is clear that the $Z_3$ and $Z_4$ survive in the
final answer because the three- and four-gluon vertices carrying the index $\mu$, 
in the diagrams $(c_1)$ and $(c_2)$ of Fig.~\ref{fig:masseq}, are bare instead of fully dressed. 
In fact, this is completely analogous to what happens in the more familiar case of the 
quark gap equation, where, by the end of the renormalization procedure, 
the quark self-energy is multiplied by the
renormalization constant $Z_1$ of the quark-gluon vertex, as shown
in the top panel of Fig.~\ref{fig:quark}.

To illustrate this correspondence in some detail, recall that the inverse of the 
full quark propagator can be written as $S^{-1}(p) =  A(p)\slashed{p} - B(p) \mathbb{I}$, 
and the dynamical quark mass function is given by \mbox{${\mathcal{M}}(p)= B(p)/A(p)$}.
In the absence of a current quark mass (chiral limit),
after the implementation of certain simplifying assumptions that do not compromise the
features we want to examine, the quark mass equation may be brought to the form~\cite{Aguilar:2010cn} 
\begin{equation}
{\mathcal{M}}(p)= c g^2_{\s R} Z_1 \int_k {\bf \rm Tr}\!\left[\Gamma_{\mu}^{(0)}S_{\s R}(k)\Gamma_{\nu}^{\s R}(-p,k,q) \Delta^{\mu\nu}_{\s R}(q)\right]\,,
\label{senergy}
\end{equation}
where $c$ is a numerical constant, the trace runs over spinor indices\footnote{The application of the trace on both sides of the gap
equation isolates the term $B(p)$.}, $q=p-k$,  
and $\Gamma_{\nu}(-p,k,q)$ denotes the fully dressed quark-gluon vertex, whose tree-level value
is given by $\Gamma_{\nu}^{(0)} = \gamma_{\nu}$.
To arrive at \1eq{senergy}, one employs the first and fourth relations in \1eq{renconst2}, together with  
\be
S_{\s R}(p) =  Z_{f}^{-1} S(p)\,,\,\,\,\,\,\,\,\,\,\,\,\, 
\Gamma_{\nu}^{\s R}(-p,k,q)  = Z_{1} \Gamma_{\nu}(-p,k,q)\,,\,\,\,\,\,\,\,\,\,\,\,\,
Z_g^{-1}  =  Z_1^{-1} Z_{f} Z_A^{1/2}\,.
\label{extraren}
\ee

The next step is to write 
the kernel of \1eq{senergy} in terms of a manifestly RGI quantity multiplied
by a momentum- and $\mu$-dependent remainder.
To that end, and in order to simplify 
the discussion, we retain only one out of the twelve tensorial structures
comprising $\Gamma_{\nu}^{\s R}(-p,k,q)$~\cite{Ball:1980ay}, namely the one proportional to its tree-level tensor, $\Gamma_{\nu}^{(0)}$.
Moreover, the form factor multiplying $\Gamma_{\nu}^{(0)}$, denoted by $L_{\s R}(-p,k,q)$, will be
evaluated in the so-called 
{\it symmetric configuration}, where $p^2=k^2=q^2$, thus becoming
a function of a single momentum~\cite{Aguilar:2018epe}, \ie 
\be
\Gamma_{\nu}^{\s R}(-p,k,q) \to L_{\s R}(q) \Gamma_{\nu}^{(0)}(-p,k,q)\,.
\label{versim}
\ee

At this point it is convenient to introduce the standard RGI quantity
\be 
{\cal R}_{\s f}(q,r,p) = g \Delta^{1/2}(q) S^{1/2}(r)S^{1/2}(p)L(q)\,,
\ee
which finally allows one to cast \1eq{senergy} into the alternative form\footnote{The trace may be carried out trivially; however, for the arguments that follow, it is advantageous 
to retain the vertices  $\Gamma^{(0)}$ manifestly in the integrand.}
\be
{\mathcal{M}}(p)= c  Z_1 \int_k L^{-1}_{\s R}(q)
{\bf \rm Tr}\!\left[\Gamma_{\mu}^{(0)} {\cal R}^2_{\s f} \Gamma_{\nu}^{(0)} {\rm P}^{\mu\nu}(q)\right]{\mathcal{M}}(k) \,,
\label{senergy2}
\ee
which is the announced result\footnote{
Note that the ratio $H(p_1)/H(p_2)$ 
of any two-point function $H(p)$ is also a RGI quantity; this fact 
may be used in order to ``compensate'' for ``mismatches'' of momenta when forming
RGI products. Such factors are immaterial for the discussion that follows and will be omitted throughout.}.

Given that ${\mathcal{M}}(p)$ is RGI, \ie $d{\mathcal{M}}(p)/d\mu=0$, the r.h.s. of \1eq{senergy2}
must display the same property; then, since ${\cal R}^2_{\s f}$ is RGI by itself,
one must have that \mbox{$d[Z_1 L^{-1}_{\s R}(q)]/d\mu=0$}. This is indeed true,
because, from the second relation in \1eq{extraren} and \1eq{versim}, 
we have \mbox{$Z_1 L^{-1}_{\s R}(q) = L^{-1}(q)$}, and, since $L(q)$ is a bare quantity, it is
trivially $\mu$-independent, \mbox{$dL(q)/d\mu=0$}.
Therefore, at this point it is clear that setting $Z_1=1$ would
distort the RG properties of the r.h.s. of \1eq{senergy}.

Evidently, the simplest way to enforce the relation \mbox{$d[Z_1 L^{-1}_{\s R}(q)]/d\mu=0$} would be to carry out
the replacement $Z_1 L^{-1}_{\s R}(q)\to {\mathcal R}$, where ${\mathcal R}$ is some RGI combination.
In fact, the most obvious ``solution'' would be to simply set ${\mathcal R}=1$, which,
interestingly enough, is precisely the one needed for recovering the correct one-loop
anomalous dimension of ${\mathcal{M}}(p)$~\cite{Aguilar:2010cn,Aguilar:2018epe}. Thus, effectively, one implements 
the substitution $Z_1\to L_{\s R}(q)$ into \1eq{senergy2}, \ie  
\be
{\mathcal{M}}(p) =  c  \int_k  {\bf \rm Tr}\!\left[\Gamma_{\mu}^{(0)}\, {\cal R}^2_{\s f} \Gamma_{\nu}^{(0)} \,{\rm P}^{\mu\nu}(q)\right] {\mathcal{M}}(k)\,.
\label{senergy3}
\ee
Clearly, due to its RGI nature, 
${\mathcal{M}}(p)$ does not depend on 
the subtraction point $\mu$ nor on the
ultraviolet cutoff $\Lambda$, \ie $d{\mathcal{M}}(p)/d\Lambda=0$.
Consequently, the implicit $\Lambda$-dependence of the integral on 
the r.h.s. of \1eq{senergy} or \1eq{senergy2}
should be canceled by the corresponding $\Lambda$-dependence of $Z_1$.
Of course, the operation $Z_1\to L_{\s R}(q)$ implemented above 
amounts to replacing a $\Lambda$-dependent constant by a $\Lambda$-independent (but $\mu$-dependent) function of $q^2$,
which, in principle, could distort the aforementioned cancellation. 
Therefore, the underlying assumption when carrying out this substitution
is that the introduction of $L_{\s R}(q)$ in the integrand of
\1eq{senergy} or \1eq{senergy2} will alter the initial $\Lambda$-dependence of the integral 
in such a way that, as $\Lambda\to\infty$, the resulting solution will satisfy 
the condition $d{\mathcal{M}}(p)/d\Lambda=0$. As we will check explicitly in 
subsection~\ref{subsec:sol}, this is indeed
what happens in the case of the gluon mass equation. 

\subsection{\label{sec:z1sde}The SDE of the quark-gluon vertex: ``solving'' for $Z_1$}
The above heuristic substitution $Z_1\to L_{\s R}(q)$ admits a simple interpretation
in the context of the SDE satisfied by the quark-gluon vertex $\Gamma^{\mu}$, being 
essentially an application of the so-called {\it dressed skeleton expansion}~\cite{Bjorken:1965zz}
(for recent treatments see, \eg ~\cite{Carrington:2012ea,Williams:2015cvx,Mueller:2015fka}).

In particular, let us consider the SDE for  $\Gamma^{\mu}$, which,
when set up from the point of view of the quark leg~\cite{Binosi:2016rxz} contains a single dressed contribution, 
shown by the diagram $(b_1)$ in Fig~\ref{fig:quark}. 
Its main ingredient is the 
amputated 4-point kernel with two gluons and a quark-antiquark pair entering in it, 
denoted by ${\cal K}_{\!\s{A\!A {\bar \psi}\!{\psi}}}$ , which is related to its renormalized counterpart,
${\cal K}_{\!\s{A\!A {\bar \psi}\!{\psi}}}^{\s R}$, by
${\cal K}_{\!\s{A\!A {\bar \psi}\!{\psi}}}^{\s R} = Z_{\s A} Z_{f} {\cal K}_{\!\s{A\!A {\bar \psi}\!{\psi}}}$.
Clearly, the combination
\mbox{$\widehat{\cal K}_{\!\s{A\!A {\bar \psi}\!{\psi}}} =\Delta\, S\, {\cal K}_{\!\s{A\!A {\bar \psi}\!{\psi}}}$}
is RGI. 
Note finally that the vertex $\Gamma^{\mu}$ appearing in graph $(b_1)$ of Fig.~\ref{fig:quark}, which is normally bare,
has been dressed up, thus converting the original SDE to its Bethe-Salpeter version;
evidently, the kernel ${\cal K}_{\!\s{A\!A {\bar \psi}\!{\psi}}}$
must be adjusted accordingly~\cite{Bjorken:1965zz,Maris:2003vk}, in order to
avoid overcounting.

Then, suppressing all indices and momenta, the SDE in the bottom panel of Fig.~\ref{fig:quark} reads
\be
\Gamma_{\!\!\s R} = Z_1 \Gamma^{(0)} + \int_\ell \Gamma_{\!\!\s R}   {\rm P} \widehat{\cal K}_{\!\s{A\!A {\bar \psi}\!{\psi}}}\,,
\ee
or, using \1eq{versim}, with appropriately assigned momenta,  
\be
Z_1 \Gamma^{(0)} = L_{\s R} \Gamma^{(0)} - \int_\ell L_{\s R} \Gamma^{(0)} {\rm P} \widehat{\cal K}_{\!\s{A\!A {\bar \psi}\!{\psi}}} \,.
\label{bjdr1}
\ee
Note the presence of a factor $L_{\s R}$ in both terms on the r.h.s. of \1eq{bjdr1}.

Then, returning  to \1eq{senergy2} and substituting 
the term $Z_1 \Gamma^{(0)}$ appearing in it by 
the r.h.s. of \1eq{bjdr1}, one obtains 
\be
{\mathcal{M}}(p) = c  \int_k {\bf \rm Tr}\!\left[\Gamma^{(0)} \,{\cal R}^2_{\s f} \,\Gamma^{(0)} \,{\rm P} \right]{\mathcal{M}}(k)
- c  \int_k \int_\ell  {\bf \rm Tr}\!\left[\Gamma^{(0)}\, \widehat{\cal K}_{\!\s{A\!A {\bar \psi}\!{\psi}}} \,{\cal R}^2_{\s f}\, \Gamma^{(0)} \,{\rm P}\,{\rm P}\right] {\mathcal{M}}(k) \,. 
\label{solvebjdr1}
\ee
Then, after neglecting the second (``higher-order'') term on the r.h.s. of \1eq{solvebjdr1}, 
one recovers precisely \1eq{senergy3}; thus, as announced, 
the above analysis boils down to the effective replacement
$Z_1 \Gamma_{\nu}^{(0)} \to L_{\s R}(q) \Gamma_{\nu}^{(0)}$.

\subsection{\label{sec:rem}Further remarks}
We point out that
the renormalization procedure adopted in~\cite{Aguilar:2018epe} is conceptually identical to the one presented above, 
but is operationally distinct, due to the use of an alternative set of approximations. In particular:
\n{i} The ghost dressing function, $F(q)$,
enters into the gap equation through the STI that $\Gamma_{\nu}$ satisfies;  
its renormalization is given by \mbox{$F_{\s R}(q)  = Z^{-1}_{c} F(q)$}.
\n{ii} In the Landau gauge, $Z_1=Z_c^{-1}$ to lowest order; the replacement of
$Z_1$ by $Z_c^{-1}$ is therefore carried out at the level of the gap equation.
\n{iii} In the Taylor renormalization scheme~\cite{Boucaud:2008gn}, the combination
 ${\mathcal R}_{\s F}(q) = g \Delta^{1/2}(q) F(q)$ is RGI.
\n{iv} By virtue of a special exact relation~\cite{Grassi:2004yq,Aguilar:2009nf},
we have $Z_c= Z_{\s G}$.

Then, the construction presented in  Sec.~\ref{sec:csb} gets modified; one considers the product 
$gZ_c^{-1}\Delta^{1/2}(q$) and converts it into a cutoff-independent RGI combination through 
replacing $Z_c^{-1}$ by a function of $q^2$.
Due to property \n{iv}, this may be accomplished in two obvious ways, namely by converting it to either ${\mathcal R}_{\s F}(q)$ or to ${\cal R}_{\s G}(q)$, which amounts to $Z_c^{-1} \to F(q)$ or $Z_c^{-1}\to[1+G]^{-1}(q)$, respectively.

In conclusion, the effective approaches of implementing multiplicative renormalizability at the
level of the quark gap equation may be summarized by the statement that one carries out the substitution 
$Z_1 \Gamma_{\nu}^{(0)} \to {\mathcal C}_1(q) \Gamma_{\nu}^{(0)}$, where, depending on the
particular details and approximations
\be
{\mathcal C}_1(q) = L_{\s R}(q)\,,\,\,\,\,\,\, {\mathcal C}_1(q) = F_{\s R}(q)\,,\,\,\,\,\,\,  {\mathcal C}_1(q) = [1+G_{\s R}(q)]^{-1}\,.
\label{threeC}
\ee
It is important to mention that all three possibilities for ${\mathcal C}_1(q)$ listed in \1eq{threeC} have the
exact same ultraviolet behavior, giving rise to the correct one-loop anomalous dimension for ${\mathcal{M}}(p)$~\cite{Aguilar:2018epe}. Quite interestingly, as may be seen in the right panel of Fig.~\ref{fig:quark}, 
these three functions are very similar in the entire range of momenta; as a result,
the solutions for ${\mathcal{M}}(p)$ obtained by inserting any one of them in the gap equation
are rather close to each other~\cite{Aguilar:2018epe}.  

\subsection{\label{sec:gmeren}Effective renormalization of the gluon mass equation}

We now return to the main objective of this section, and model the multiplicative
renormalization of the gluon mass equation following a method completely analogous to the one outlined above. 

To begin with, let us point out that, unlike ${\mathcal{M}}(p)$, the $m^2(q)$ is {\it not} RGI.
Nonetheless, the quark construction may be followed closely, by simply introducing,
for the purposes of this discussion, the 
dimensionless RGI quantity ${\overline m}^2(q) :=  m^2(q)/m^2(0)$. 
Then \1eq{fullmass} remains the same, except
for the substitutions $m^2(q)\to {\overline m}^2(q)$
and $m^2(k)\to {\overline m}^2(k)$ on its l.h.s and r.h.s., respectively, which are trivially implemented after dividing both sides by the (nonvanishing) $m^2(0)$.

To proceed further, let us consider $\Gamma_{\!3}(q_1,q_2,q_3)$ and $\Gamma_{\!4}(p_1,p_2,p_3,p_4)$,
and simplify their structures, in a way analogous to what was done 
in \1eq{versim} for $\Gamma_{\nu}^{\s R}(-p,k,q)$. To that end, 
consider a single form factor for each vertex, proportional to their 
corresponding tree-level structures, namely 
\be
\Gamma_{\!3}(q_1,q_2,q_3) \rightarrow {\cal C}_3(s)  \Gamma^{(0)}_{\!3}(q_1,q_2,q_3)\,,\,\,\,\,\,\,\,\,\,\,\,\,  
\Gamma_{\!4}(p_1,p_2,p_3,p_4) \rightarrow {\cal C}_4 (s) \Gamma^{(0)}_{\!4}(p_1,p_2,p_3,p_4)\,,
\label{c3c4}
\ee
where we have, at the symmetric points\footnote{Two scales $s_1$ and $s_2$ may be chosen, instead
  of the common $s$, without affecting the central argument.}, 
$q_1^2 =q_2^2=q_3^2 := s^2$ and
$p_1^2 =p_2^2=p_3^2=p_4^2 := s^2$, respectively\footnote{The corresponding inner products are given by 
$q_i\cdot q_j = - s/2$ and $q_i\cdot q_j = - s/3$ ($i \neq j$), respectively \cite{Gracey:2014ola}.}.

In addition, we introduce the following two RGI combinations~\cite{Aguilar:2014tka}, 
\be 
{\cal R}_3(s) =  g \Delta^{3/2}(s)\, {\cal C}_3(s),
\,\,\,\,\,\,\,\,
{\cal R}_4(s)=  g^2  \Delta^{2}(s)\, {\cal C}_4(s)\,,
\label{R3R4a}
\ee
which, due to the particular kinematics chosen, depend only on a single variable $s$. 

Then, the two strings appearing on the r.h.s. of \1eq{renstrings} may be re-expressed as  
\bea
Z_3\, g^2_{\s R} \,\Delta^{2}_{\s R} \,[1+G_{\s R}(q)]^{-1} &=& Z_3\, {\cal C}^{-1}_{3\s R} \,{\cal R}_{\s G} {\cal R}_3\,,
\nonumber\\
Z_4\, g^4_{\s R}\, \Delta^{4}_{\s R} \,{\cal C}_{3\s R} \, [1+G_{\s R}(q)]^{-1} &=& Z_4 \,{\cal C}^{-1}_{4\s R} \, {\cal R}_{\s G} {\cal R}_3  {\cal R}_4 \,.
\label{R3R4}
\eea
Therefore, one may rewrite \1eq{fullmass} in the following schematic form (suppressing irrelevant kinematic factors) 
\be
{\overline m}^2(q) \sim \int_k {\overline m}^2(k) \left\{
Z_3\, {\cal C}^{-1}_{3\s R} \,{\cal R}_{\s G} {\cal R}_3 + Z_4 \,{\cal C}^{-1}_{4\s R} \, {\cal R}_{\s G} {\cal R}_3  {\cal R}_4
\right\}\,,
\label{fm2} 
\ee
which is the analogue of \1eq{senergy2}. Then, following essentially the same reasoning,
one implements the substitutions
$Z_3\to {\cal C}_{3\s R}$  and  $Z_4 \to {\cal C}_{4\s R}$, or, equivalently, setting $s=k$,  
\bea
{\cal K}^{+}_{\s R}(q,k) &\rightarrow & {\cal K}^{+}_{\rm{\s{eff}}}(q,k) = {\cal C}_4(k) [\Y(k+q) + \Y(k)] - {\cal C}_3(k)\,,
\nonumber\\
{\cal K}^{-}_{\s R}(q,k) &\rightarrow & {\cal K}^{-}_{\rm{\s{eff}}}(q,k) = {\cal C}_4(k) [\Y(k+q) - \Y(k)]\,.  
\label{renkernel2}
\eea
This final step is depicted in panel $(D)$ of Fig.~\ref{fig:masseq}.


\subsection{\label{sec:z3z4sde}``Solving'' for $Z_3$ and $Z_4$ from the vertex SDEs}

\begin{figure}[t]
\begin{minipage}[b]{1.0\linewidth}
\centering
\includegraphics[scale=0.4]{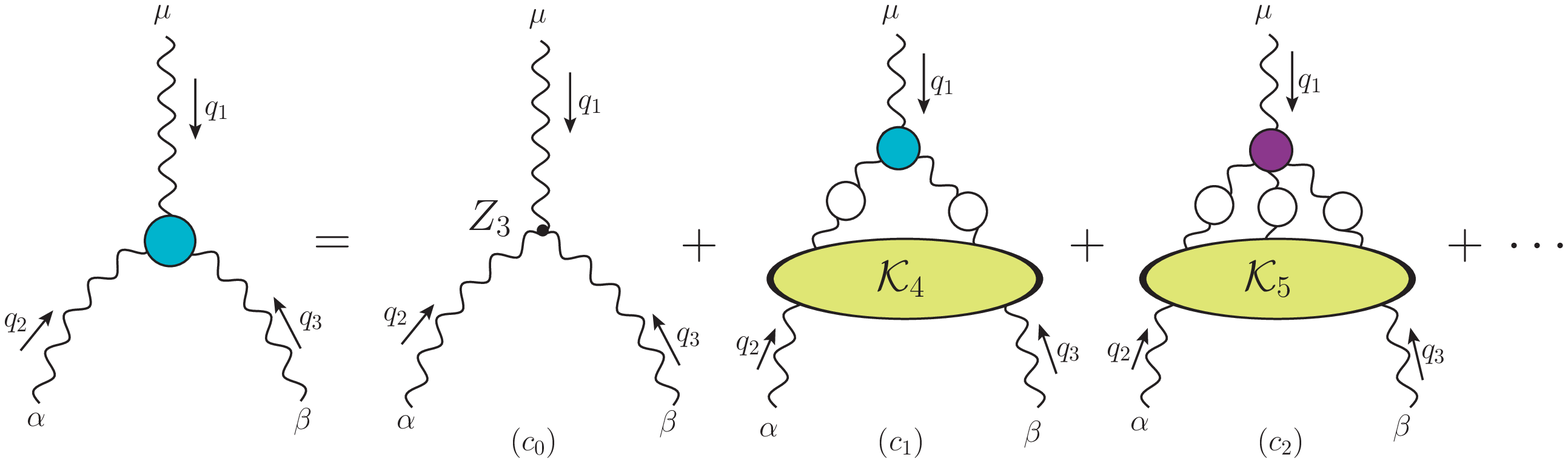} 
\vspace{0.5cm}
\end{minipage}
\begin{minipage}[b]{1.0\linewidth}
\includegraphics[scale=0.4]{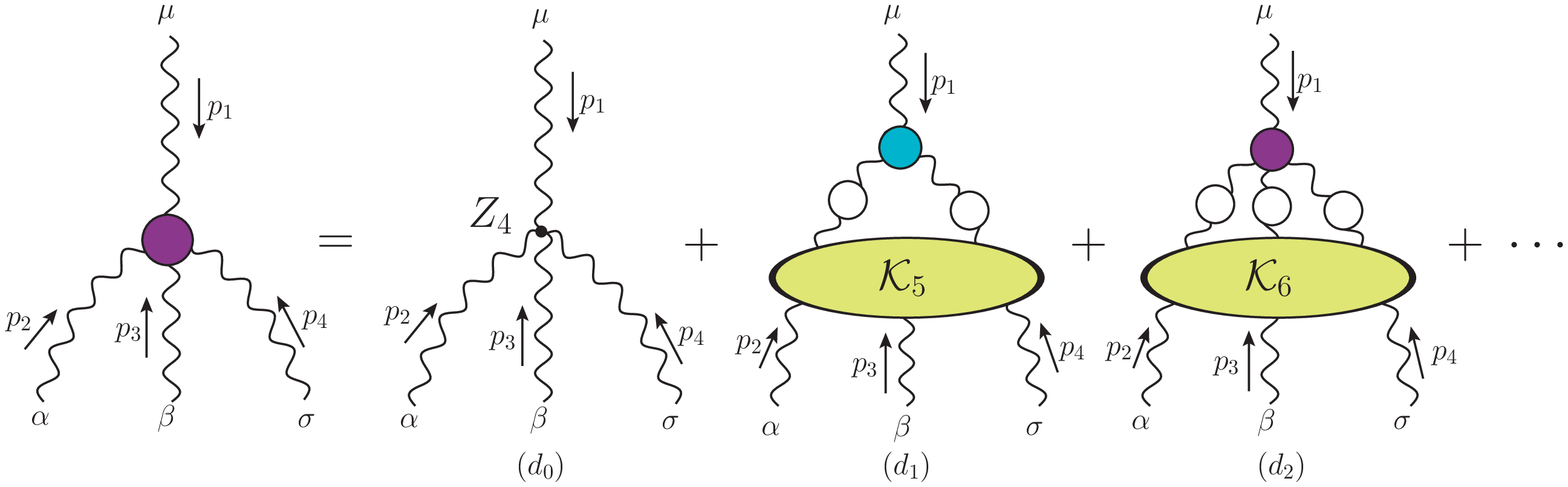}
\end{minipage}
\caption{The SDEs for the three-gluon and four-gluon vertices.}
\label{fig:vertices}
\end{figure}

The construction presented in Sec.~\ref{sec:z1sde} may be repeated for the case in hand,
by considering the SDEs for the vertices $\Gamma_{\!3}$ and $\Gamma_{\!4}$, 
represented in Fig.~\ref{fig:vertices}, whose main ingredients 
are multi-gluon kernels.
In particular, suppressing color and Lorentz indices, 
we denote  by  ${\cal K}_{n}$ the amputated kernels with $n$ incoming gluons,
and  by ${\cal K}_{n}^{\s R}$ their renormalized counterparts;
the kernels are related to each other by
${\cal K}_{n}^{\s R} = Z^{\frac{n}{2}}_{\s A} \, {\cal K}_{n}$. 
Then, the combinations \mbox{$\widehat{\cal K}_{n} =  \Delta^{\!\frac{n}{2}} {\cal K}_{n}$},
are clearly RGI.

To illustrate these definitions with an example, consider 
the ``lowest'' order dressed contribution to ${\cal K}_{4}$, to be denoted by ${\cal K'}_{\!\!4}$, given by
\be
{\cal K'}_{\!\!4} = g^2 \Gamma_{\!3} \Delta \Gamma_{\!3} = Z_g^2 Z_3^{-2} Z_{\s A}
\{ g^2_{\s R} \Gamma_{\!3}^{\s R}  \Delta^{\!\!\s R}\Gamma_{\!3}^{\s R}\}   = Z_{\s A}^{-2} {\cal K'}_{\!\!4}^{\s R} \,.
\ee
Then, the $\widehat{\cal K}'_{4}$ is 
given by $\widehat{\cal K}'_{4} = \Delta^2 \{g^2 \Gamma_3 \Delta\Gamma_{3}\} = {\cal R}^2_3$.

Turning to the SDEs, and suppressing strings of projectors ${\rm P}$ that are totally inert, we have
\bea
\Gamma_{\!3}  &=& Z_3 \Gamma^{(0)}_{\!3} + \int_k \Gamma_{3} \widehat{\cal K}_{4} \, +  \int_k\int_\ell  \widehat{\cal K}_{5} \{g \Delta^{1/2} \Gamma_{4}\} \,\,+ ...\,,
\nonumber\\
\Gamma_{\!4}  &=& Z_4 \Gamma^{(0)}_{\!4} 
+ \int_k \widehat{\cal K}_{5}  \{g^{-1} \Delta^{-1/2} \Gamma_{3}\}\,
+ \int_k\int_\ell  \Gamma_{4} \widehat{\cal K}_{6} \,\, + ...\,,
\label{versdes}
\eea
where the ellipses denote the remaining terms of Fig.~\ref{fig:vertices}.

Then, from \1eq{R3R4}, we have that
$g \Delta^{1/2} {\cal C}_4 = {\cal C}_3 {\cal R}_4 {\cal R}^{-1}_3$, or, equivalently,  
\mbox{$g^{-1} \Delta^{-1/2} {\cal C}_3 = {\cal C}_4 {\cal R}_3 {\cal R}^{-1}_4$}.
Substituting appropriately,  \1eq{versdes} may be expressed as 
\bea
{\cal C}_3 \Gamma^{(0)}_{\!3}  &=& Z_3 \Gamma^{(0)}_{\!3} + \int_k {\cal C}_3 \Gamma^{(0)}_{\!3} \widehat{\cal K}_{4} +
\int_k \int_\ell {\cal C}_3 \Gamma^{(0)}_{\!4} {\cal R}_4 {\cal R}^{-1}_3 \widehat{\cal K}_{5} \,\,+ ...\,,
\nonumber\\
    {\cal C}_4 \Gamma^{(0)}_{\!4}  &=& Z_4 \Gamma^{(0)}_{\!4}
    + \int_k {\cal C}_4 \Gamma^{(0)}_{\!3} {\cal R}_3 {\cal R}^{-1}_4 \widehat{\cal K}_{5}
   \, + \int_k\int_\ell  {\cal C}_4 \Gamma^{(0)}_{\!4} \widehat{\cal K}_{6} \,\, + ...\,.
\label{versdes2}
\eea
Evidently, after this rearrangement, the first equation in \1eq{versdes2} involves only ${\cal C}_3$, 
while the second only ${\cal C}_4$. Thus, the relations analogous to \1eq{bjdr1} become
\bea
Z_3 \Gamma^{(0)}_{\!3}  &=& {\cal C}_3 \Gamma^{(0)}_{\!3} -
\int_k {\cal C}_3 \left[\Gamma^{(0)}_{\!3} \widehat{\cal K}_{4} + \int_\ell\Gamma^{(0)}_{\!4} {\cal R}_4 {\cal R}^{-1}_3 \widehat{\cal K}_{5}\right]\,\,- ...\,,
\nonumber\\
Z_4 \Gamma^{(0)}_{\!4}  &=& {\cal C}_4 \Gamma^{(0)}_{\!4} - \int_k {\cal C}_4 \left[\Gamma^{(0)}_{\!3} {\cal R}_3 {\cal R}^{-1}_4 \widehat{\cal K}_{5} \,
+\int_\ell \Gamma^{(0)}_{\!4} \widehat{\cal K}_{6} \right]- ...\,.
\label{bjdr2}
\eea
Clearly, and in exact analogy with \1eq{bjdr1}, 
the omission of the integral contribution of the r.h.s. of
both equations leads to the announced heuristic substitution 
$Z_3\Gamma^{(0)}_{\!3} \to {\cal C}_3 \Gamma^{(0)}_{\!3}$ and $Z_4\Gamma^{(0)}_{\!4} \to {\cal C}_4 \Gamma^{(0)}_{\!4}$.

\section{\label{sec:inp} The main ingredients of the numerical analysis}

In this section, we first cast the mass equation into a form appropriate for its numerical treatment,
and subsequently discuss the main characteristics and physical properties of the ingredients entering in it.

In order to solve \1eq{fullmass} numerically, we switch to spherical coordinates, 
introducing the variables $x=q^2$, $y=k^2$, and $z= (k+q)^2=x+y+2\sqrt{xy} c_{\theta}$, 
where $c_{\theta} := \cos\theta$, $s_{\theta} := \sin\theta $, and 
\be
\int_k = \frac{1}{(2\pi)^3} \int_{y,\theta}\,,\,\,\,\,\,\,\,\,
\int_{y,\theta} := \int_0^\infty\! dy\, y \int_0^\pi\! d\theta\,s_{\theta}^{2} \,.  
\label{sphercord}
\ee
Then, the equation to solve assumes the form 
\be
m^2(x)= \frac{\alpha_s C_A}{2\pi^2}\frac{1}{x\,[1+G(x)]}
\int_{y,\theta} z^{-1} \Delta(y)\Delta(z) [{\cal K}_1(x,y,z) + {\cal K}_2(x,y,z)]m^2(y) \,,
\label{eumass}
\ee
where
\bea
{\cal K}_1(x,y,z) &=& \left\{ {\cal C}_4(y)\left[\Y(z)+\Y(y)\right]- {\cal C}_3(y)\right\} (z-y) \left(3z - x s_{\theta}^{2}\right) \,,
\nonumber\\
{\cal K}_2(x,y,z) &=& {\cal C}_4(y) \left[\Y(z)-\Y(y) \right]\left[x(z + y s_{\theta}^{2}) +2 (z-y)^2\right] \,.
\label{newKs}
\eea
As already mentioned in the Introduction,
in all previous works \1eq{eumass} has been linearized, by treating 
the $\Delta(y)$ and $\Delta(z)$ as external inputs, whose form was determined from appropriate fits to the gluon 
lattice data of~\cite{Bogolubsky:2007ud}.
Instead, in the present analysis
we maintain the nonlinear nature of \1eq{eumass} intact, by replacing~\1eq{eq:gluon_m_J} in it, \ie setting
\be
\Delta(t) = [t J(t) + m^2(t)]^{-1} \,, \qquad t= y,z\,.
\label{nonlinD}
\ee

We next discuss the main characteristics and physical properties of the
various ingredients entering in \1eq{eumass}, and in particular of 
$J(q)$, ${\cal C}_3(k)$, ${\cal C}_4(k)$,  $Y(k)$, and $1+G(q)$.

{\it (i)}
In order to implement \1eq{nonlinD}, and in the absence of a {\it bona fide} dynamical equation,
a suitable {\it Ansatz} for $J(q)$ needs to  be employed, which will be gradually
improved during the iterative procedure (see next section). 

\begin{figure}[t]
\begin{minipage}[b]{0.45\linewidth}
\centering 
\hspace{-1.5cm}
\includegraphics[scale=0.27]{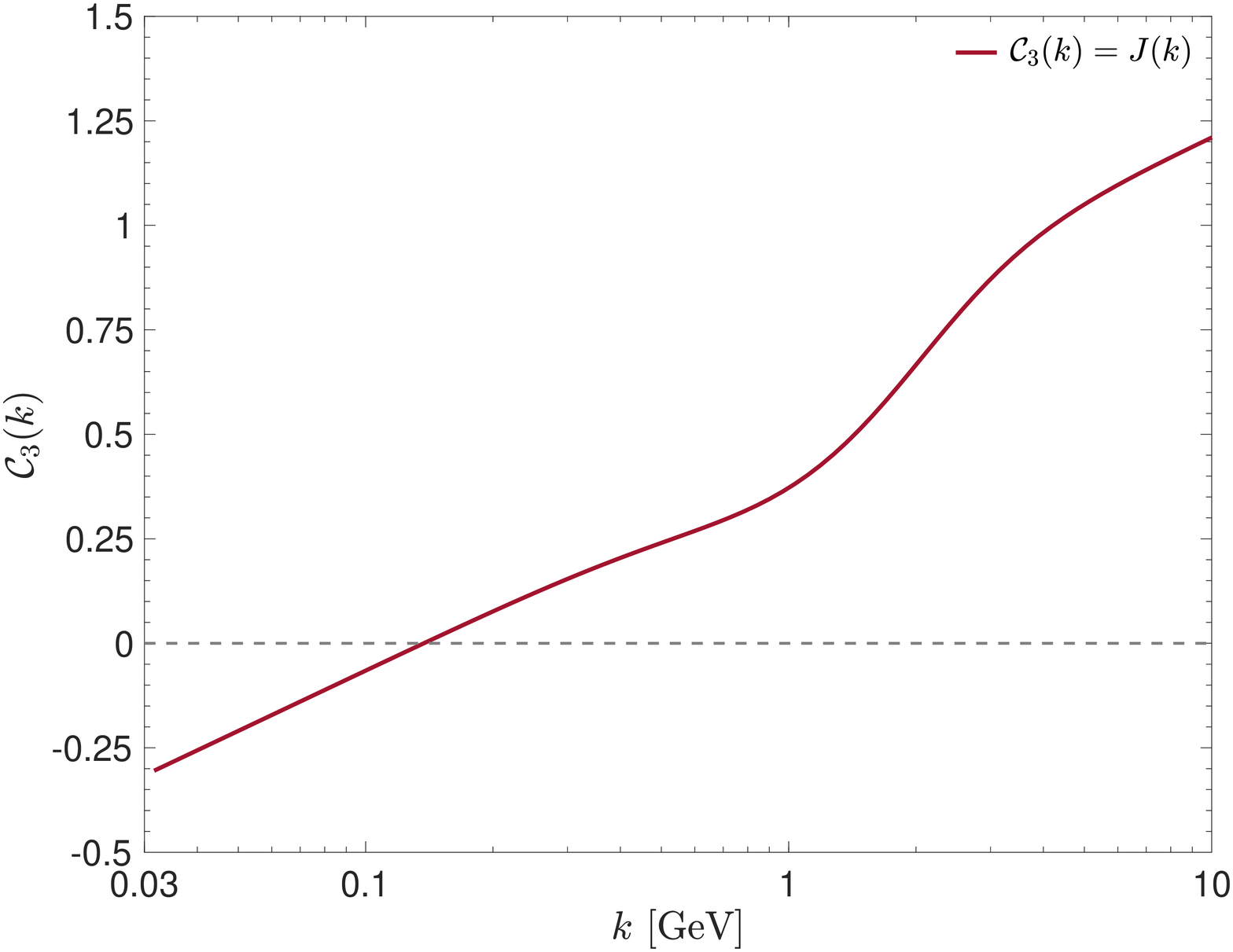}
\end{minipage}
\hspace{0.15cm}
\begin{minipage}[b]{0.45\linewidth}
\includegraphics[scale=0.27]{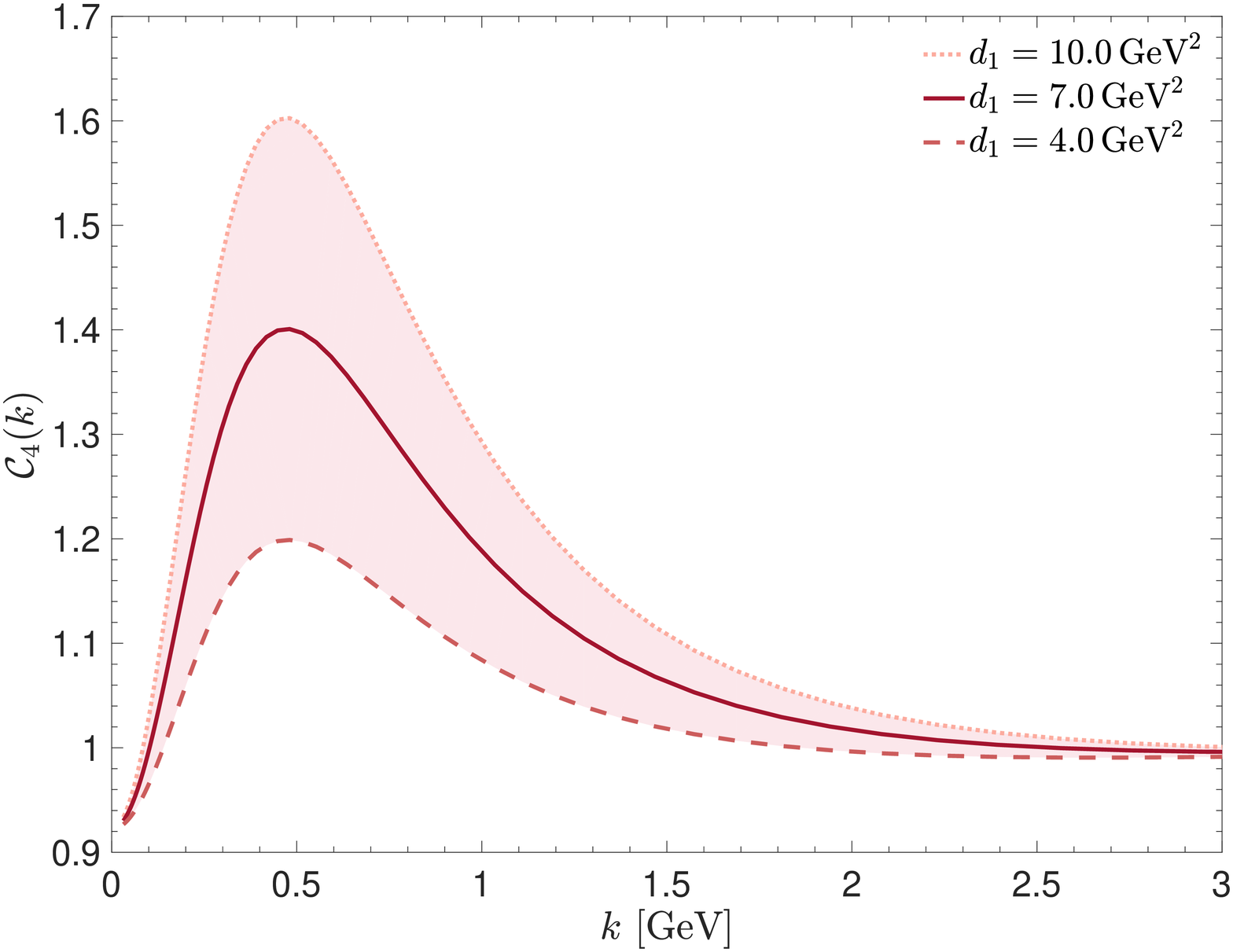}
\end{minipage}
\caption{Left panel: The ${\cal C}_3(k)$ employed in the first iteration. Right panel: The ${\cal C}_4(k)$ given by Eq.~\eqref{c4}, with the parameter $d_1$ varying in the range \mbox{$(4.0-10.0)\,\mbox{GeV}^2$}.}
\label{fig:C34}
\end{figure}

In the left panel of Fig.~\ref{fig:C34} we show the initial seed for $J(q)\to J_0(q)$; it displays the same functional form employed in the recent nonperturbative Ball-Chiu construction of the longitudinal
part of the three gluon vertex~\cite{Aguilar:2019jsj}, namely
\be  
\label{eq:J_fits} 
J(q) = 1 + \frac{ C_\mathrm{A} \lambda_s }{ 4 \pi }\left( 1 +
\frac{ \tau_1 }{ q^2 + \tau_2 } \right) \left[ 2 \ln\left( \frac{q^2 + \eta^2(q) }{\mu^2} \right) + \frac{1}{6}\ln\left( \frac{q^2}{\mu^2} \right) \right] \,,
\ee
with
\be 
\eta^2(q) = \frac{\eta_1}{q^2+ \eta_2} \,,
\label{eta}
\ee
where the fitting parameters for $J_0(q)$ are quoted in Table~\ref{tableJ}. It is important to emphasize that, throughout this work, the renormalization point will be fixed at \mbox{$\mu=\mbox{4.3}$ GeV}. 

{\it (ii)} According to its definition in Eq.~\eqref{c3c4}, ${\cal C}_3(s)$  
is the co-factor of the tree-level structure of $\Gamma_{\!3}(q_1,q_2,q_3)$  
when all form-factors are evaluated at the symmetric point.
In particular, in the Ball-Chiu basis (see Eq.~\eqref{eq:3g_sti_structure} and Eq.~(3.4) of~\cite{Aguilar:2019jsj}), 
we have that, at the symmetric point, $X_1(s)= X_4(s)= X_7(s) := {\cal C}_3(s)$.
In general, the $X_i(q,r,p)$ may be expressed in terms of $J(q)$, the ghost dressing function
$F(q)$, and three of the five form factors of the so-called ghost-gluon kernel~\cite{Aguilar:2018csq}.
However, the corresponding nonperturbative evaluation 
reveals that the ``Abelian approximation'', obtained by turning off the ghost sector,
is numerically rather close to the full answer (see, \eg Fig.~7 in~\cite{Aguilar:2019jsj}).
Therefore, we will simplify the complexity of our analysis by using the corresponding
``Abelian'' result (see Eq.~(3.13) of~\cite{Aguilar:2019jsj}), \eg
\be
{\cal C}_3(k) = J(k)\,,
\label{C3def}  
\ee
which is represented in the left panel of Fig.~\ref{fig:C34}.
Evidently, since the form of $J(k)$ will vary from one iteration to the next, by virtue of \1eq{C3def} 
so will ${\cal C}_3(k)$.

{\it(ii)} 
Unfortunately, the available functional studies~\cite{Binosi:2014kka,Cyrol:2014kca,Cyrol:2016tym,Huber:2018ned} furnish rather limited information  
on the nonperturbative properties of the four-gluon vertex, and no lattice simulations have been carried out to date\footnote{See also~\cite{Gracey:2014ola,Gracey:2019mix} for a variety of relevant properties of the four-gluon vertex.}. 
Therefore, our {\it Ansatz} for ${\cal C}_4(k)$ will be designed to simply
capture certain general trends, observed in all aforementioned studies.
In particular, for a variety of special kinematic configurations, described by a single momentum scale, 
the form factor accompanying either the $\Gamma^{(0)}_{\!4}$ or its {\it transversely projected counterpart} 
displays a typical peak, located in the region of a few hundred MeV. 
Motivated by the above observations, the overall qualitative behavior of ${\cal C}_4(k)$ will be modeled by
\be
{\cal C}_4(k)= 1 +\frac{\lambda}{4\pi}\left[1 - \frac{d_1k^2}{(k^2+d_2)^2}\right]\ln\left(\frac{k^2+4m^2_0}{\mu^2}\right)\,,
\label{c4}
\ee
where
\mbox{$\lambda=0.28$}, \mbox{$d_2=0.26\,\mbox{GeV}^2$}, and \mbox{$m^2_0=0.14\,\mbox{GeV}^2$}; the corresponding curves are shown on the right panel of Fig.~\ref{fig:C34}. Notice that the red shaded area is created varying the value of $d_1$  in the range of \mbox{$(4.0-10.0)\,\mbox{GeV}^2$}, while all other parameters in Eq.~\eqref{c4} are kept fixed. 

As we will see in the end of Sec.~\ref{sec:numan}, these variations of ${\cal C}_4(k)$  have no appreciable impact on our solutions, and
may be compensated by appropriately re-adjusting the value of $\alpha_s$.
The aspect that seems to be decisive is the moderate enhancement that ${\cal C}_4(k)$ displays with respect to its tree-level value (unity) in a region of momenta known to be important for mass generation.

\begin{figure}[t]
\begin{minipage}[b]{0.45\linewidth}
\centering 
\hspace{-1.5cm}
\includegraphics[scale=0.27]{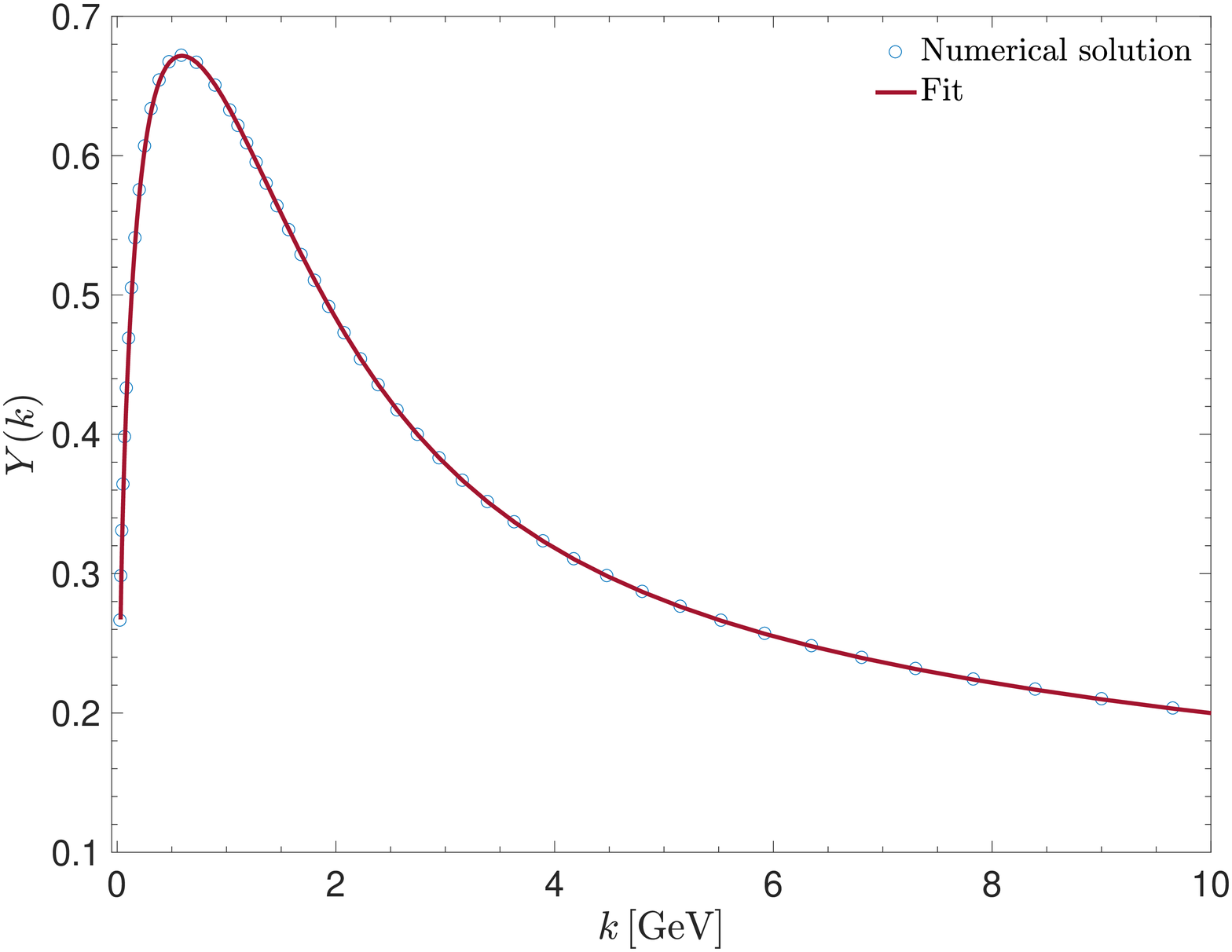}
\end{minipage}
\hspace{0.15cm}
\begin{minipage}[b]{0.45\linewidth}
\includegraphics[scale=0.27]{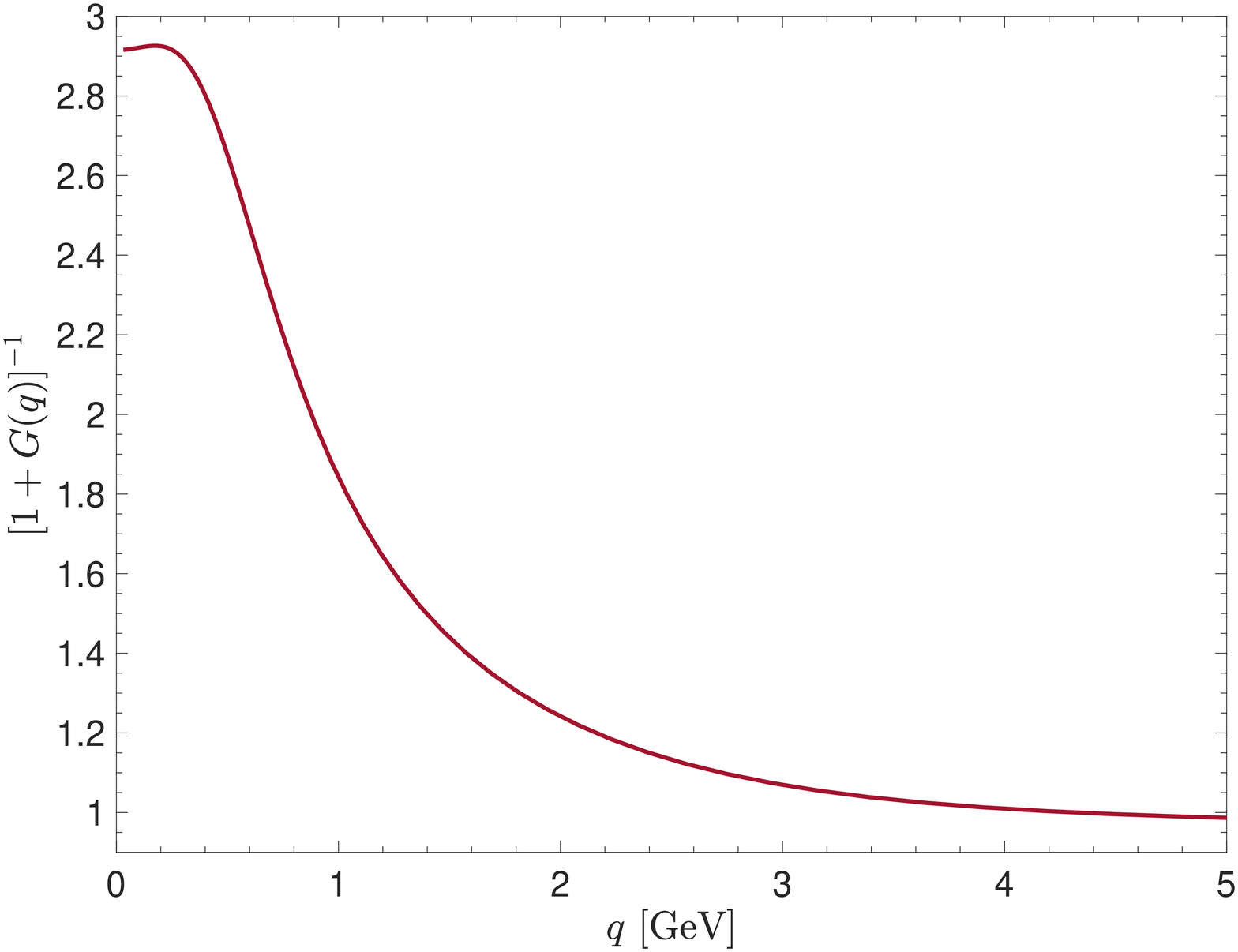}
\end{minipage}
\caption{Left panel: The numerical solution for $Y(k)$ obtained from~\1eq{Ysph} (circles), and the corresponding fit given by Eq.~\eqref{fitY} (continuous). Right panel:  The inverse of the 
 auxiliary function, $1+G(q)$, whose fit is given by Eq.~\eqref{fit1G}.}
\label{fig:Y}
\end{figure}

{\it (iv)} The determination of $Y(k)$ proceeds by evaluating numerically its defining expression~\cite{Binosi:2012sj}
\be
\Y(k)= -\frac{1}{4} g^2C_A \frac{k^\rho}{k^2}\int_\ell\Delta_{\mu\rho}(\ell)\Delta_{\alpha\nu}(\ell+k)\Gamma_{\!3}^{\alpha\mu\nu}(k,\ell,-k-\ell)\,.
\label{Ydef}
\ee
To that end, we set $\Gamma_{\!3}^{\alpha\mu\nu} = \Gamma_{\!3\s{\mathbf{L}}}^{\alpha\mu\nu} + \Gamma_{\!3\s{\mathbf{T}}}^{\alpha\mu\nu}$, 
where $\Gamma_{\!3\s{\mathbf{L}}}^{\alpha\mu\nu}$ saturates the relevant STIs, 
while the $\GT^{\alpha\mu\nu}$ vanishes when contracted by $q_{\alpha}$, $r_{\mu}$, or $p_{\nu}$. Keeping only the former term, 
we have that  
\be
\Gamma_{\!3\s{\mathbf{L}}}^{\alpha\mu\nu}(q,r,p) = \sum_{i=1}^{10} X_i(q,r,p) \ell_i^{\alpha\mu\nu} \,,
\label{eq:3g_sti_structure}
\ee
where the basis tensors $\ell_i^{\alpha\mu\nu}$ are given in Eq.~(3.4) of~\cite{Aguilar:2019jsj}.
After carrying out the various momentum contractions,
and passing to spherical coordinates, one arrives at
\be 
\Y(y) = \frac{\alpha_s C_A}{8\pi^2} \int_{t,\omega} s^2_{\omega} \Delta(t)\Delta(u) \mathcal{K}_{\s \Y}(t,y,\omega) \,,
\label{Ysph}
\ee
where $y = k^2$, $t = \ell^2$, $u = (k + \ell)^2 = y + t + 2 \sqrt{yt} c_{\omega}$, 
\be
\mathcal{K}_{\s \Y}(t,y,\omega)= - t X_6 + 6 X_7 -(u+y-t)\left[3X_9 + \frac{t}{u}X_3 \right]+ \frac{(u+t-y)}{2u}
\left[ X_4 - 2 X_1 \right]\,,
\label{KXi}
\ee
and $X_i = X_i(y,t,\omega)$. Note that the additional $\sin^2\!\omega$ in the angular integral stems from the  
presence of the common factor $\frac{k^2\ell^2-(k\cdot \ell)^2}{k^2 \ell^2} = s^2_{\omega}$.

To further evaluate $\Y(y)$ through \2eqs{Ysph}{KXi}, we employ the results for the form factors $X_i$
obtained in~\cite{Aguilar:2019jsj}\footnote{In earlier works, $Y(k)$ was determined either 
by setting $\Gamma_{\mu\alpha\beta}(q,r,p)=\Gamma_{\mu\alpha\beta}^{(0)}(q,r,p)$~\cite{Binosi:2012sj,Aguilar:2014tka},
or by using the first relation in \1eq{c3c4},
where the functional form of ${\cal C}_3(s)$, denoted by $f(s)$ in~\cite{Binosi:2017rwj},
is given by Eq.~(5.5) of that article.}. 
The curve obtained  
is shown in the left panel of Fig.~\ref{fig:Y}; it can be fitted by
\be
Y(k) =3\pi\alpha_{\s s} C_{\rm A}\left\{\left[A\ln\left(\frac{k^2+\eta^2(k)}{\mu^2}\right)
+B\ln\left(\frac{k^2}{\mu^2}\right)
\right]\left[1+ \frac{Ck}{1+(k^2/\nu^2)^{\gamma}}\right] + D \right\}  \,,
\label{fitY} 
\ee
where $\eta^2(k)$ is given by Eq.~\eqref{eta}.  The fitting parameters are 
\mbox{$A=-0.015$}, \mbox{$B=0.0095$}, \mbox{$C=2.158$ GeV},  \mbox{$D=0.039$}, \mbox{$\nu^2=2.422\,\mbox{GeV}^2$}, \mbox{$\gamma=1.074$}, \mbox{$\eta_1=0.0103\,\mbox{GeV}^4$}, and \mbox{$\eta_2=0.184\,\mbox{GeV}^2$}.  
As we will see in the next section, the concrete value of $\alpha_{\s s}$ will be tuned, for each set of ingredients,
at the level of the dynamical equation; the curve shown in the left panel of Fig.~\ref{fig:Y}
is obtained by using $\alpha_{\s s} =0.27$.

{\it (v)}  The final ingredient is the auxiliary function $1+G(q)$,
introduced in \1eq{BQIs}, 
whose inverse is shown in the right panel of Fig.~\ref{fig:Y}. For 
this function we employ the following fit, which is valid for the entire range of Euclidean momenta~\cite{Aguilar:2018epe},
namely 
\be
1+G(q) = 1 + \frac{9C_{\rm A}}{48\pi}[\alpha_{g} + A_1\exp{\left(-\rho_1 q^2/\mu^2\right)} ]\ln\left(\frac{q^2+\rho_2\eta^2(q)}{\mu^2}\right)\,,
\label{fit1G}
\ee
where $\eta^2(q)$ is also given by Eq.~\eqref{eta}, but now with \mbox{$\eta_1=0.30\,\mbox{GeV}^4$}, \mbox{$\eta_2=0.33\,\mbox{GeV}^2$}. The remaining adjustable parameters are  
\mbox{$\alpha_g=0.21$}, \mbox{$A_1=0.77\,\mbox{GeV}^2$}, \mbox{$\rho_1=0.78$}, and \mbox{$\rho_2=0.50$}.

\section{\label{sec:numan} Solutions of the nonlinear mass equation}

Having defined all necessary inputs, in this section we discuss in detail the solutions obtained from the
numerical treatment of the gluon mass equation.

\subsection{\label{subsec:gqo} General qualitative observations} 

Before embarking on the full analysis, we address certain qualitative issues related with this particular equation.

We start by observing that, as  $x\to 0$,  Eq.~\eqref{eumass} reduces itself to the following  nontrivial constraint 
\be
m^2(0)=-\frac{3C_A\alpha_s}{8\pi}[1+G(0)]^{-1}\int_0^\infty\!\!\!\!\! dy\, m^2(y)\mathcal{K}_0(y)\,,
\label{mass0}
\ee
where 
\begin{align}
\mathcal{K}_0(y)=\mathcal{C}_3(y)\left[y^2\Delta^2(y)\right]^\prime- 2\, \mathcal{C}_4(y)\left[y^2\Delta^2(y)Y(y)\right]^\prime\,.
\label{K0}
\end{align}
Note that, when \mbox{$\mathcal{C}_3(y)=\mathcal{C}_4(y)=1$},
Eq.~\eqref{mass0} collapses to Eq.~(8.11) of~\cite{Binosi:2012sj}\footnote{We emphasize that, for convenience, the definition of $Y(k)$ in Eq.~\eqref{newKs}  {\it absorbs}
  a factor \mbox{$C =3\pi\alpha_s C_{\rm A}$},  
which in~\cite{Binosi:2012sj} multiplies explicitly the $Y$ terms.}. 
Eq.~\eqref{mass0} is especially useful, because it captures in a relatively simple expression
some of the crucial features displayed by the full equation.

\begin{figure}[t]
\begin{minipage}[b]{0.45\linewidth}
\centering 
\hspace{-1.5cm}
\includegraphics[scale=0.27]{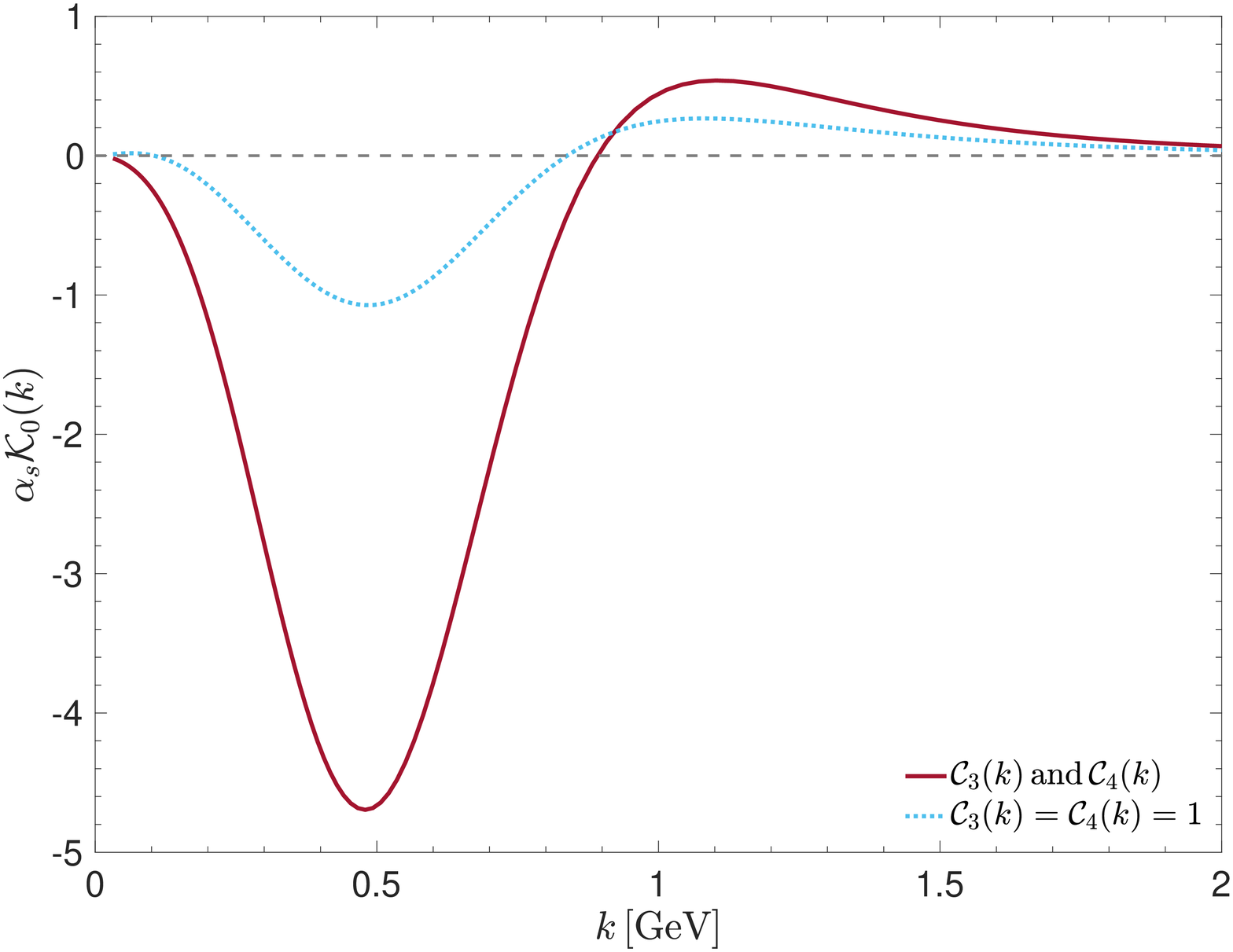}
\end{minipage}
\hspace{0.15cm}
\begin{minipage}[b]{0.45\linewidth}
\includegraphics[scale=0.27]{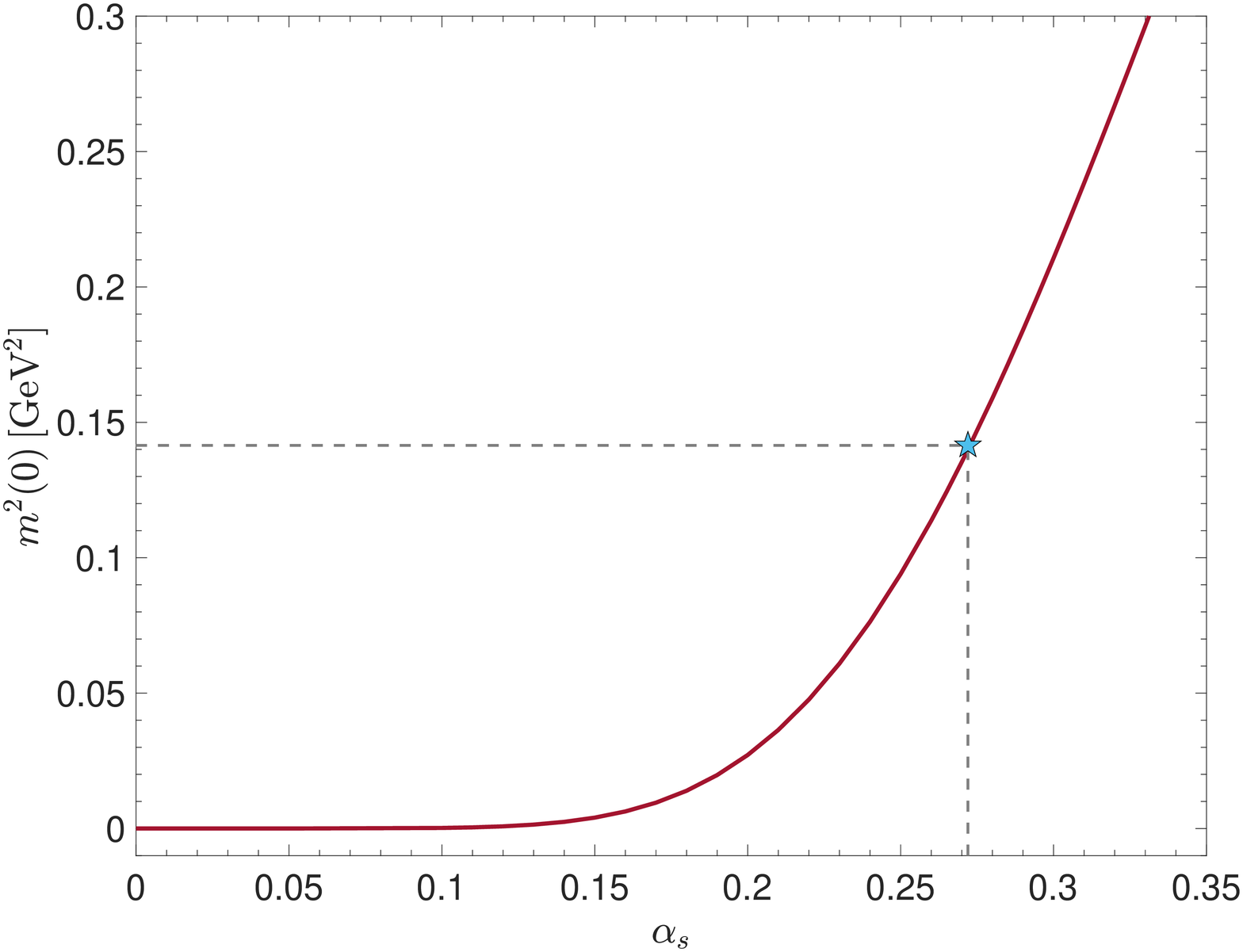}
\end{minipage}
\caption{Left panel: The kernel $\alpha_s\mathcal{K}_0(k)$ defined by Eq.~\eqref{K0} for {\it (i)}
  $\mathcal{C}_3(y)$ and $\mathcal{C}_4(y)$ given by Eqs.~\eqref{C3def} and~\eqref{c4},
respectively (red continuous), and {\it (ii)} for \mbox{$\mathcal{C}_3(y)=\mathcal{C}_4(y)=1$} (blue dotted).
For both cases we have used $\alpha_s =0.27$.
Right panel: The values of $m^2(0)$, obtained from solving \1eq{eumass} for a \emph{fixed} $J(q)$, 
as function of $\alpha_s$. The blue star denotes the $\alpha_s$ that reproduces the lattice value $m^2(0)=0.14$.} 
\label{fig:const}
\end{figure}

We start by highlighting the impact that ${\mathcal C}_3(k)$ and ${\mathcal C}_4(k)$
have on the structure of the kernel~\eqref{K0}. Specifically, the net effect of both functions
is to broaden considerably 
the negative support of the kernel 
with respect to the case \mbox{$\mathcal{C}_3(y)=\mathcal{C}_4(y)=1$} [see left panel of Fig.~\ref{fig:const}]; 
consequently, the equation may accommodate comfortably a positive-definite $m^2(y)$.

Furthermore, it is important to emphasize from the outset that,
contrary to what happens in the linearized case~\cite{Binosi:2012sj,Aguilar:2014tka},
where solutions exist only for a unique value of $\alpha_s$, the nonlinearized equation
yields solutions for a continuous (and rather extended) interval of values for $\alpha_s$.
The simplest way to establish this is to vary $\alpha_s$ keeping the form of  $J(q)$ fixed,
and observe that one obtains a continuous family of $m^2(q)$ [see right panel of Fig.~\ref{fig:const}].
Of course, the $m^2(q)$ so obtained, when put together with the  $J(q)$ in the combination of \1eq{eq:gluon_m_J},
give rise to gluon propagators that, in general, have little or nothing to do with the lattice results for $\Delta(q)$.
As we will see below, in order to approach the lattice data, the values of $\alpha_s$ must be chosen from a rather narrow interval.

 \subsection{\label{subsec:sol}Full numerical analysis: results and discussion}

{\it The numerical procedure:} The numerical solution for $m^2(q)$ is obtained through an iterative procedure
consisting of the following main steps:

($s_0$) An excellent numerical fit to the gluon lattice data of~\cite{Bogolubsky:2007ud} is employed, to be
denoted by $\Delta_{\s {\rm L}}(q)$; its functional form is given in Eq.~(4.1) of~\cite{Aguilar:2010cn}.
In particular, we fix the fitting parameters such that \mbox{$\Delta^{-1}_{\s {\rm L}}(0) = 0.14\,\mbox{GeV}^2$}.

($s_1$): We begin the iteration by introducing two initial seeds, one for $m^2(q)$ and another one for $J(q)$.
 For $m^2(q)$ we use a random function, while for $J(q)$ the {\it Ansatz} of \1eq{eq:J_fits}, \ie we set 
$J(q) \to J_0(q)$; the corresponding fitting parameters are quoted in Table~\ref{tableJ}. 

($s_2$): With these starting ingredients, we solve 
 \1eq{eumass} iteratively, adjusting the value of $\alpha_s$ such that $m^2(0)=\Delta^{-1}_{\s {\rm L}}(0)$.
The solution is accepted when the relative difference between two successive results for $m^2(q)$ 
is below $10^{-5}$; we denote this solution by $m^2_{s_2}(q)$. 

($s_3$): The  $m^2_{s_2}(q)$ is combined with the $J_0(q)$ as dictated by \1eq{eq:gluon_m_J},
in order to obtain our approximation for $\Delta(q)$, which is then compared with $\Delta_{\s {\rm L}}(q)$.

($s_4$): In order to improve the result of ($s_3$), we determine a new $J(q)$, which will be used
to obtain from \1eq{eumass} a new solution for $m^2(q)$. This new  $J(q)$ is obtained from  \1eq{eq:gluon_m_J}, 
\ie $J(q) = [\Delta^{-1}_{\s {\rm L}}(q) - m^2_{s_2}(q)]/q^2$. The resulting $J(q)$ is fed into \1eq{eumass},
and the step ($s_2$) repeated. 

($s_5$): The steps ($s_2$)-($s_4$) are repeated, saving those combinations of $m^2(q)$ and $J(q)$ which best 
reproduce $\Delta_{\rm L}(q)$.

\begin{figure}[!ht]
\begin{minipage}[b]{0.45\linewidth}
\centering
\hspace{-1.5cm}
\includegraphics[scale=0.27]{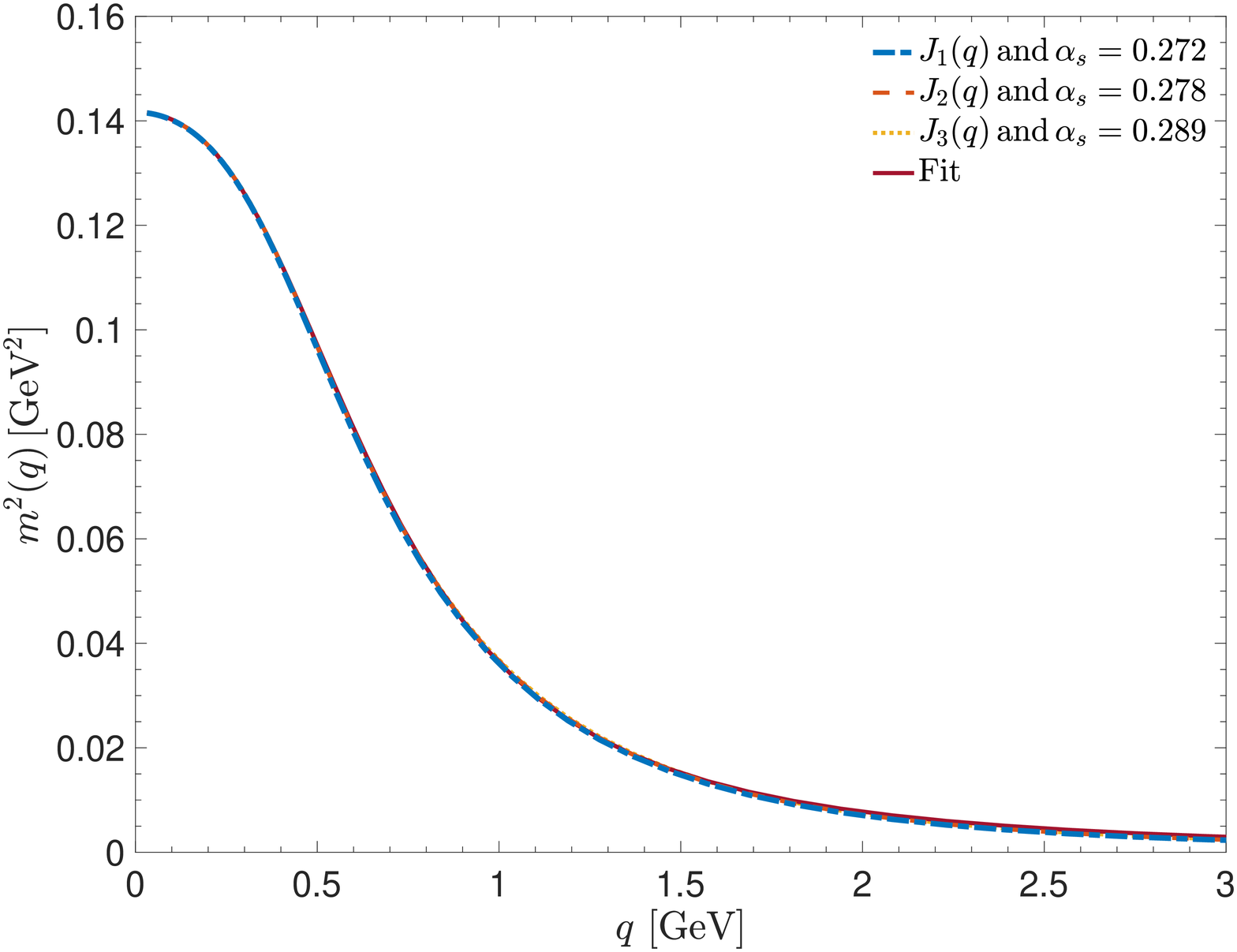}
\end{minipage}
\hspace{0.15cm}
\begin{minipage}[b]{0.45\linewidth}
\includegraphics[scale=0.27]{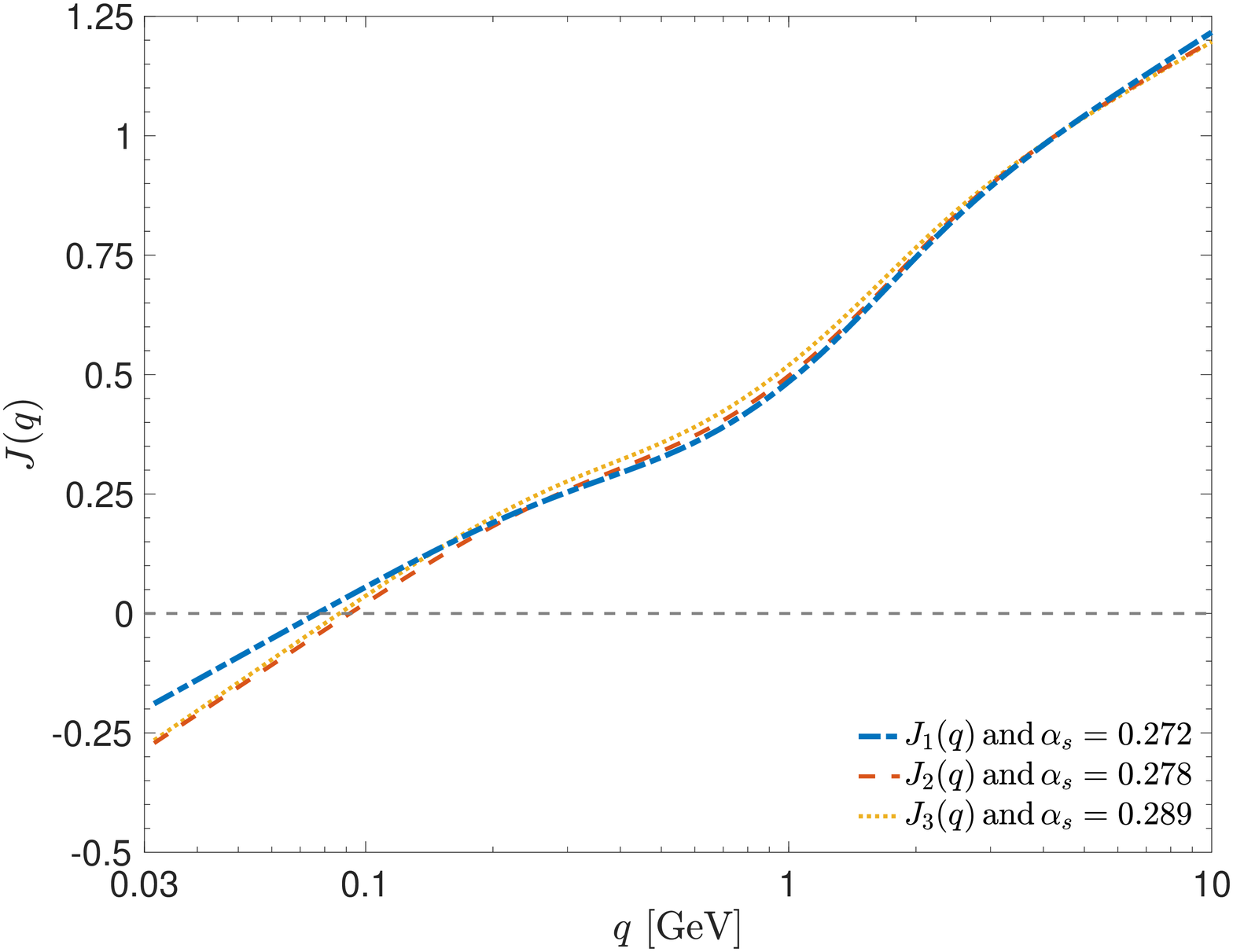}
\end{minipage}
\begin{minipage}[b]{0.45\linewidth}
\includegraphics[scale=0.27]{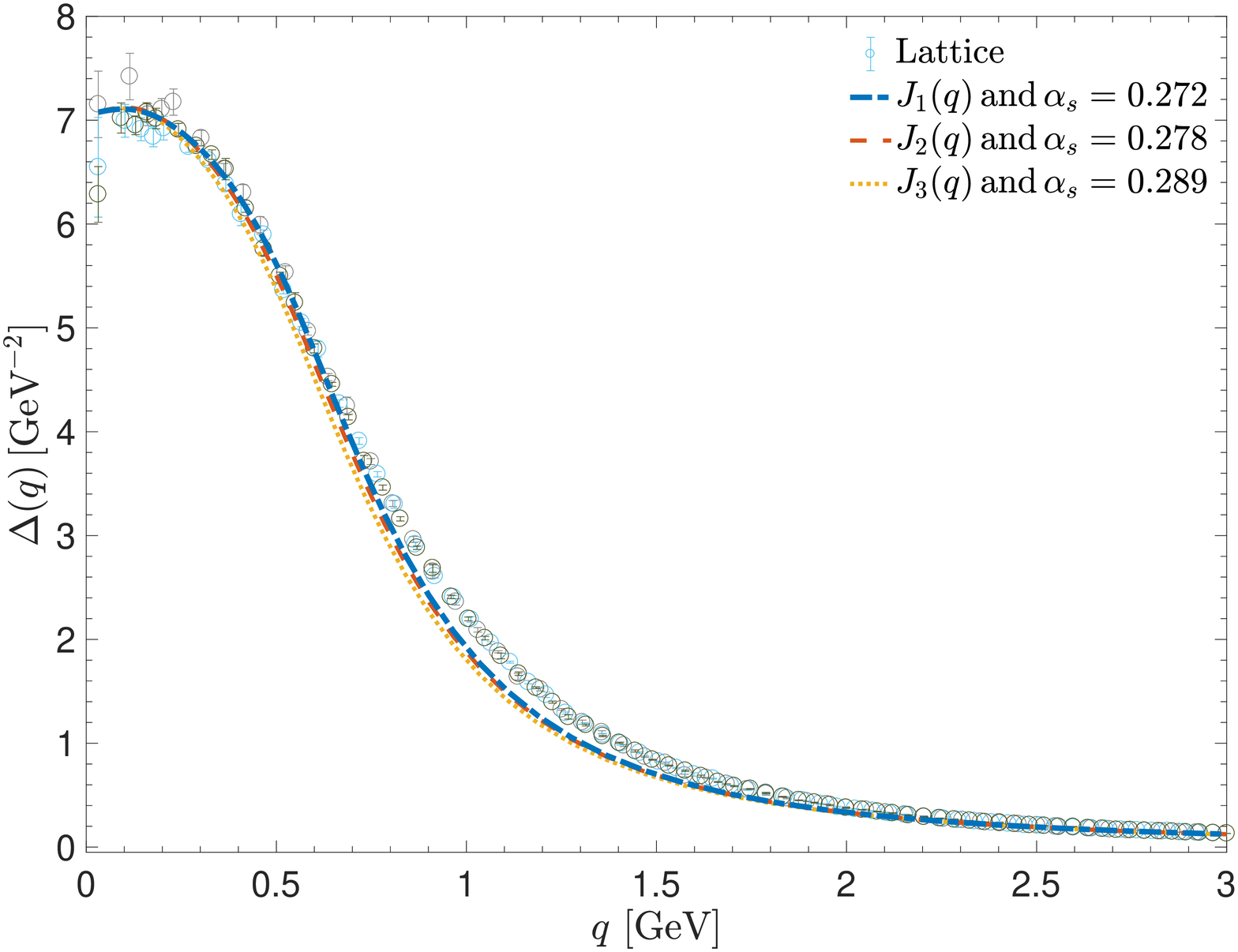}
\end{minipage}
\caption{ Top left panel: The numerical results for the  dynamical gluon mass $m^2(q)$, for $\alpha_s=0.272$ (blue dashed dotted), $\alpha_s=0.278$ (red dashed), and $\alpha_s=0.289$ (yellow dotted). Top right panel: The corresponding kinetic term $J(q)$.  Bottom panel: 
The resulting gluon propagator $\Delta(q)$ obtained from \1eq{eq:gluon_m_J}. The lattice data is from~\cite{Bogolubsky:2007ud}. In all plots,
we employ the same color code.}
\label{fig:solution}
\end{figure}

In Fig.~\ref{fig:solution} we present the outcome of the procedure described above, for three different cases of $J(q)$.
In particular, we show the best results obtained for each case, which occur 
when $\alpha_s=0.272$ (blue dashed dotted curves), $\alpha_s=0.278$ (red dashed), and 
$\alpha_s=0.289$ (yellow dotted). As mentioned at step ($s_2$) above, these values of 
$\alpha_s$ are essentially determined from the requirement that
\mbox{$\Delta_{\s {\rm L}}(0) = m^{-2}(0)$}. Evidently, this condition is
rather restrictive, forcing  $\alpha_s$ to take values within a rather small interval, {\it i.e. \mbox{$\alpha_s \in [0.272,0.289]$}}, with the renormalization point fixed at $\mu=4.3$ GeV.
Quite interestingly, this range is completely compatible with the analysis of~\cite{Boucaud:2008gn},
and is particularly close to $\alpha_s=0.32$, which is the estimated value of the coupling used in 
the lattice simulations of~\cite{Athenodorou:2016oyh,Boucaud:2017obn}.

It becomes clear from the top panels of Fig.~\ref{fig:solution}, that small variations in the $J(q)$ can be compensated by minor adjustments in the value of $\alpha_s$, producing basically the same solution for $m^2(q)$. 

In what follows, we will comment on the main characteristics of each plot
shown in Fig.~\ref{fig:solution} and their subsequent applications.

{\it (i)} We start with the \emph{dynamical gluon mass},  $m^2(q)$, shown 
in the top left panel.  As one can clearly see, $m^2(q)$ is positive-definite and monotonically
decreasing in the entire range of momenta. In addition, it may be accurately fitted  with the  characteristic power-law running  given by  
\be
m^2(q) = \frac{m_0^4}{\mu_1^2 + q^2\ln\left[(q^2+ \mu_2^2)/{\lambda^2}\right]} \,,
\label{massfit}
\ee
where the fitting parameters are fixed at
\mbox{$m_0^4=0.107\,\mbox{GeV}^4$}, \mbox{$\mu_1^2=0.756\,\mbox{GeV}^2$}, \mbox{$\mu_2^2=0.266\,\mbox{GeV}^2$}, and  \mbox{$\lambda^2=0.123\,\mbox{GeV}^2$}.

\begin{figure}[!ht]
\begin{minipage}[b]{0.45\linewidth}
\centering 
\hspace{-1.5cm}
\includegraphics[scale=0.27]{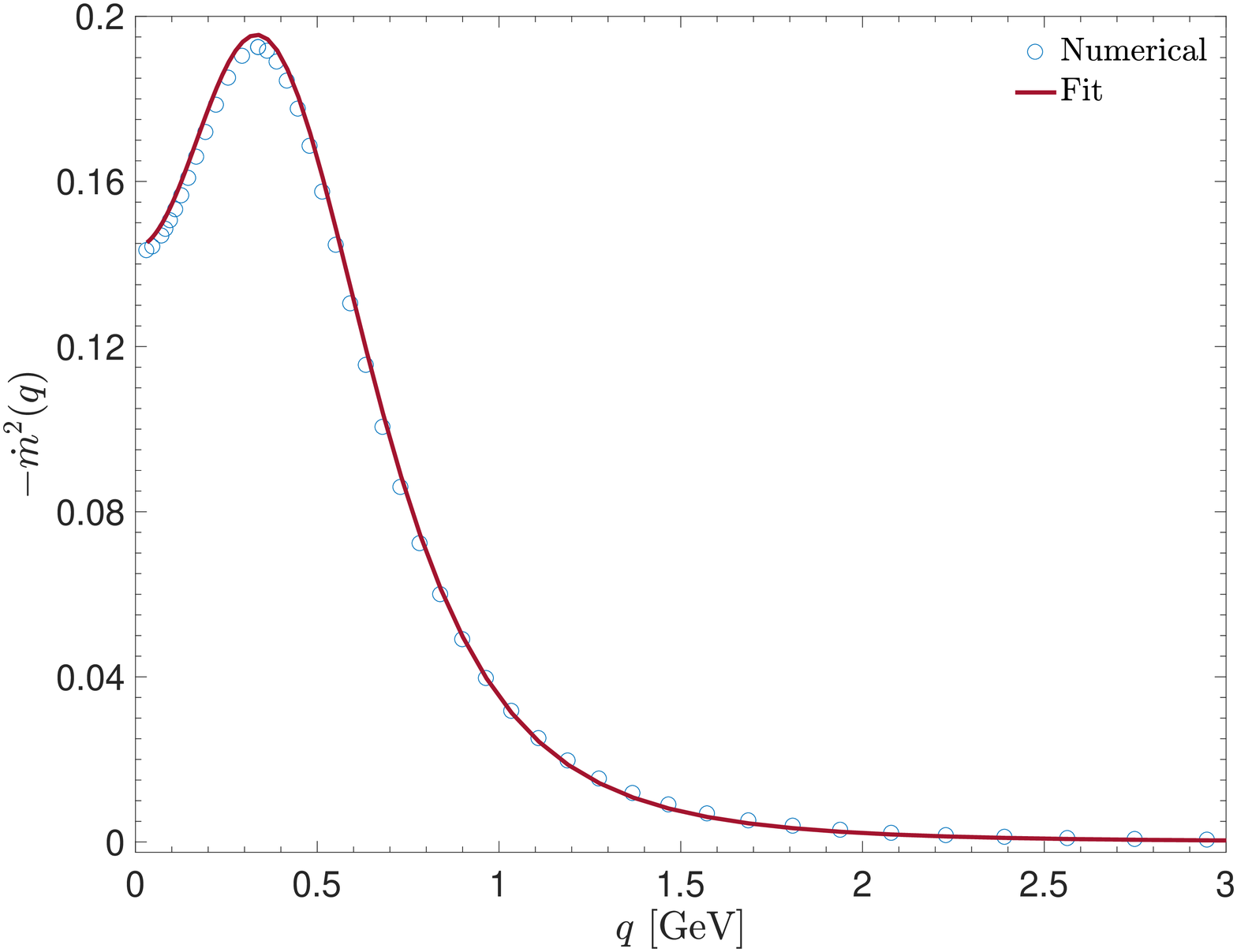}
\end{minipage}
\hspace{0.15cm}
\begin{minipage}[b]{0.45\linewidth}
\includegraphics[scale=0.27]{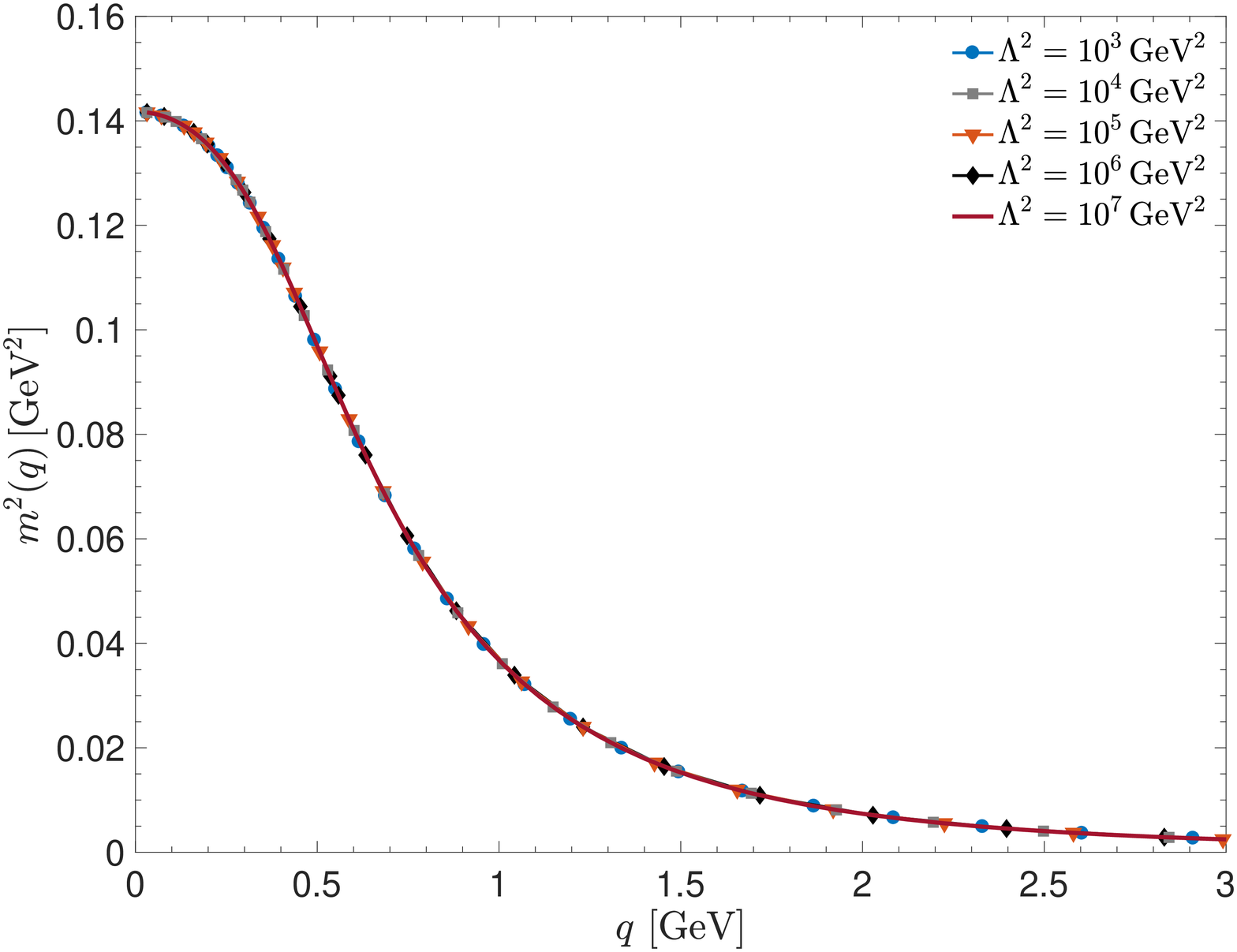}
\end{minipage}
\caption{Left panel: Comparison of the quantity $-{\dot m}^2(q)$ obtained from differentiating 
 {\it (i)} the numerical data (blue circles), and {\it (ii)}
  the fit given in \1eq{massfit} (red continuous curve). Right panel: The cutoff-independence of the
numerical solution for $m^2(q)$.}
\label{fig:massB}
\end{figure}

We emphasize that this particular fit is superior to previous
ones put forth in the related literature~\cite{Aguilar:2015bud,Aguilar:2019jsj},
(\eg $m^2(q^2) =  m^2_0/[1+(q^2/\lambda^2)^{1 + \gamma}], \,\gamma>0 $),  
because it captures faithfully not only $m^2(q)$, but also its first derivative with respect to $q^2$,
to be denoted by ${\dot m}^2(q)$.
In particular, as we can verify in the left panel of Fig.~\ref{fig:massB}, the result of the differentiation of the fit
in \1eq{massfit} is practically identical to the numerical differentiation of the ``raw'' data 
for $m^2(q)$. In fact, one may easily establish that the aforementioned sub-optimal fit  
yields a derivative that vanishes at the origin, a feature which is certainly not
shared by the actual numerical solution.
The importance of reproducing correctly this derivative is related 
to the fact that the quantity $-{\dot m}^2(q)$ is {\it exactly} equal to the Bethe-Salpeter amplitude
that controls the formation of the massless excitation that triggers the
Schwinger mechanism, and the subsequent generation of a gluon mass~\cite{Aguilar:2011xe,Binosi:2017rwj}
[see also the related discussion in Sec.~\ref{sec:conc}].

In addition, as stated in Sec.~\ref{sec:csb}, 
in the right panel of Fig.~\ref{fig:massB} we show that $m^2(q)$ is independent of the ultraviolet cutoff $\Lambda^2$, introduced for the numerical evaluation of the ``radial'' part of \1eq{sphercord}. 
Specifically, we vary $\Lambda^2$ in the range of \mbox{$(10^3-10^7)\,\mbox{GeV}^2$},
and we clearly observe that all curves lie on top of each other. 

{\it (ii)} The \emph{kinetic term} $J(q)$ is shown in the top right panel of Fig.~\ref{fig:solution},
for three values of $\alpha_s$. 
Evidently, the three curves $J_i(q)$ ($i=1,2,3$) are mild variations of 
the initial {\it Ansatz} $J_0(q)$ shown in Fig.~\ref{fig:C34}; their 
differences are related with the location of the zero crossing,
which is shifted towards lower momenta with respect to $J_0(q)$, 
and the ``bending'' displayed in the intermediate region. In particular, 
the zero crossings are located at 
\mbox{$q=78$ MeV} (blue dashed dotted), \mbox{$q=96$ MeV} (red dashed), and \mbox{$q=90$ MeV} (yellow dotted). 
We note that the $J_i(q)$ may also be fitted by the same functional form as 
the initial {\it Ansatz} $J_0(q)$, namely Eq.~\eqref{eq:J_fits}; the corresponding    
fitting parameters for the three cases are quoted in Table~\ref{tableJ}.

\begin{table}[ht]
\begin{tabular}{|c|c|c|c|c|c|}
\hline
\hline
$J(q)$ &\;$\lambda_s$\;&$\tau_1\,[\mbox{GeV}^2]$ & $\tau_2\,[\mbox{GeV}^2]$ &$\eta_1\,[\mbox{GeV}^4]$& $\eta_2\,[\mbox{GeV}^2]$\\ 
 \hline
\hline
$J_0(q)$&$0.220$ &$9.870$ &$0.910$ &$17.480$ & $1.180$ \\
$J_1(q)$&$0.243$ &$2.638$ &$0.265$ &$6.451$ & $0.388$ \\
$J_2(q)$&$0.220$ &$3.503$ &$0.263$ & $8.261$ & $0.454$ \\
$J_3(q)$&$0.220$ &$2.8$ &$0.201$ &$6.489$ & $0.363$ \\
\hline
\hline
 \end{tabular}
  \vspace{0.25cm}
\caption{The fitting parameters for $J_i(q)$ whose functional form is given by Eq.~\eqref{eq:J_fits}. 
$J_0(q)$ is the initial {\it Ansatz} presented in Fig.~\ref{fig:C34}, while $J_1(q)$,  $J_2(q)$, and 
$J_3(q)$ are the solutions shown in the top right panel of Fig.~\ref{fig:solution}.}
 \label{tableJ}    
 \end{table}

An interesting check of the overall quality of the $J_i(q)$
shown above may be obtained by means of the connections established
in~\cite{Aguilar:2019jsj}. As was explained there,
the nonperturbative generalization of the Ball-Chiu construction~\cite{Ball:1980ax}
allows one to express the ``longitudinal'' form factors of the
three-gluon vertex $\Gamma_{\!3}^{\mu\alpha\beta}(q,r,p)$ in terms of 
the kinetic term  $J(q)$ and three of the components of the so-called
ghost-gluon kernel~\cite{Aguilar:2018csq}. The form factors 
so obtained may be then used  
to estimate some of the quantities measured in lattice simulations
of the three-gluon vertex~\cite{Athenodorou:2016oyh}. One typical such quantity, denoted by 
$L^{\rm sym}(Q)$, involves a special combination of vertex form factors evaluated
at the symmetric point (\mbox{$q^2=r^2=p^2=Q^2$}); for its exact definition, see
~\cite{Athenodorou:2016oyh,Aguilar:2019jsj}.

\begin{figure}[!ht]
\includegraphics[scale=0.27]{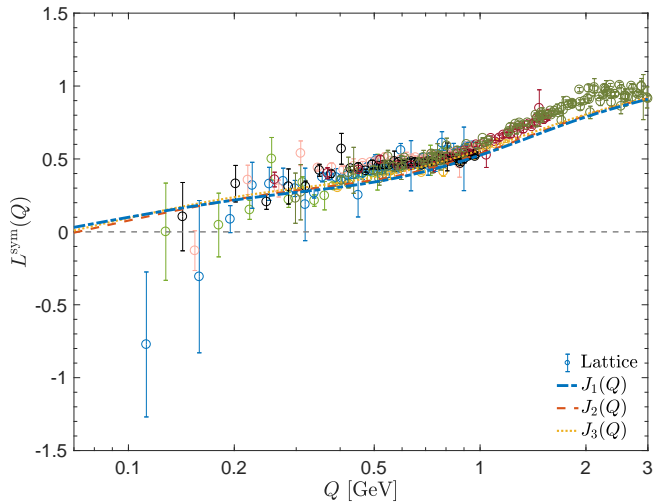}
\caption{Comparison between the lattice data of~\cite{Athenodorou:2016oyh} and 
the $L^{\rm sym}(Q)$ obtained using as input the $J_i(q)$ of Fig.~\ref{fig:solution}.}
\label{fig:lsym}
\end{figure}

In Fig.~\ref{fig:lsym}, we compare the lattice data of~\cite{Athenodorou:2016oyh} with 
the results for $L^{\rm sym}(Q)$ obtained by substituting the $J_i(q)$ of Fig.~\ref{fig:solution} into the 
Ball-Chiu solution given in Eq.~(3.11) of~\cite{Aguilar:2019jsj};  
evidently, the general shape of the lattice data is reproduced rather accurately.
Note that, since the iteration procedure shifts the 
zero-crossing of each $J_i(q)$ towards the infrared, 
the corresponding zero-crossing of $L^{\rm sym}(Q)$ display the same tendency, being at \mbox{59 MeV}, \mbox{76 MeV}, and \mbox{70 MeV}, respectively.
This result is to be contrasted 
with the left panel of Fig.~16 in~\cite{Aguilar:2019jsj}, where the predicted zero-crossing of $L^{\rm sym}(Q)$  
occurs at higher momenta \mbox{($109-155$) MeV}.

{\it (iii)} The comparison of our results for the \emph{gluon propagator}, $\Delta(q)$, with
the lattice data of~\cite{Bogolubsky:2007ud} is shown in the bottom panel of Fig.~\ref{fig:solution}; 
one can see that the pairs,  $J_i(q)$ and $m^2(q)$, reproduce rather well the lattice
data in the entire range of momenta.
Notice that the largest discrepancy  between our calculated $\Delta(q)$ and the lattice data occurs  
in the region of momenta between \mbox{$(0.8-2.5)$ GeV}, where the  relative
error  ranges from $[0.1-0.15]$ for $J_1(q)$ (blue dashed dotted curve), $[0.1-0.16]$ for $J_2(q)$ (red dashed), and $[0.1-0.2]$ for $J_3(q)$ (yellow dotted curve).  For lower momenta,  the relative errors drop considerably, becoming of the order of $10^{-2}$.
Clearly, the intermediate region is more sensitive to the truncations and approximations implemented; 
nonetheless, it is quite notable that our solution for $\Delta(q)$ reproduces very well the entire momenta range, 
by appropriately tuning the value of $\alpha_s$. 

\begin{figure}[t]
\begin{minipage}[b]{0.45\linewidth}
\centering 
\hspace{-1.5cm}
\includegraphics[scale=0.27]{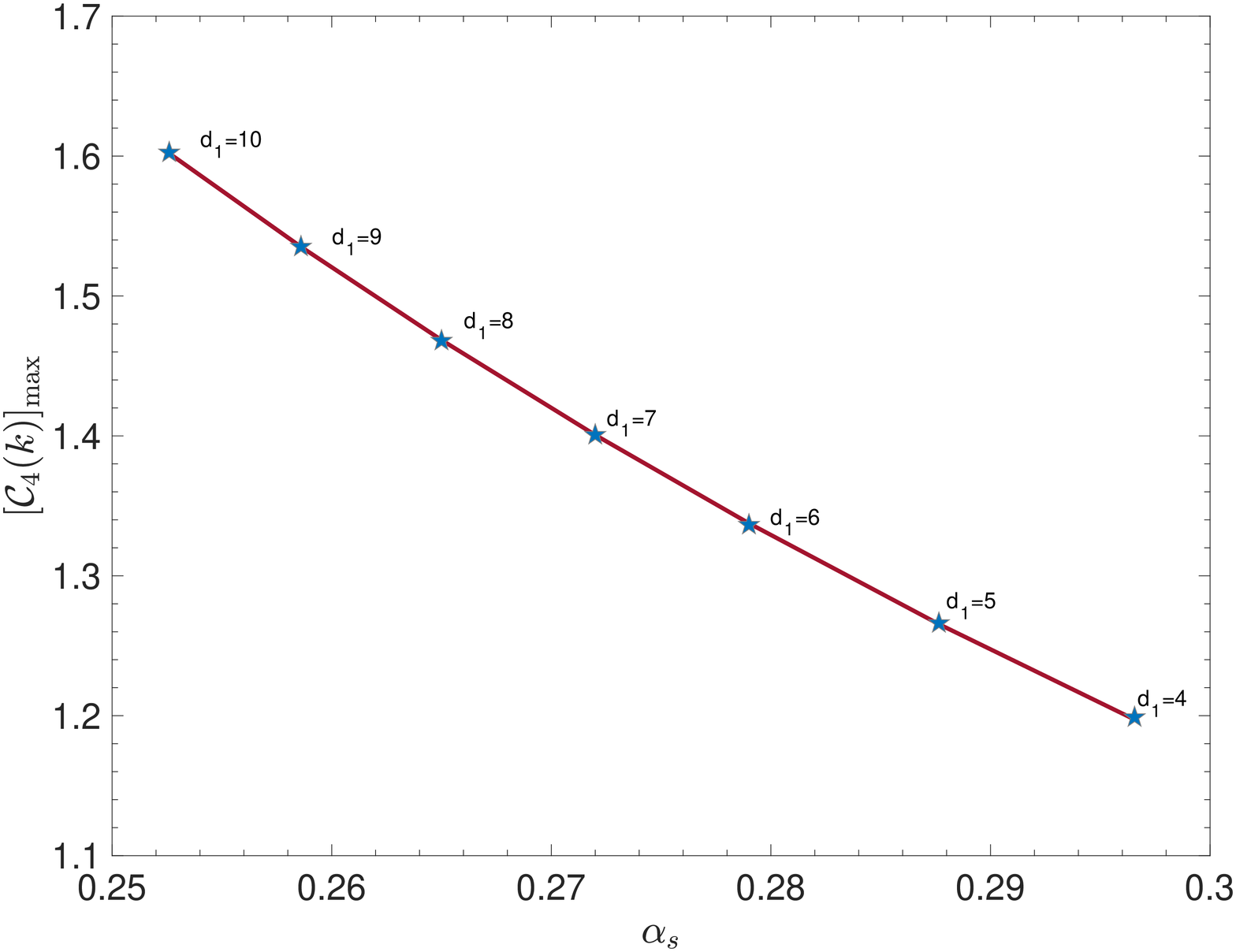}
\end{minipage}
\hspace{0.15cm}
\begin{minipage}[b]{0.45\linewidth}
\includegraphics[scale=0.27]{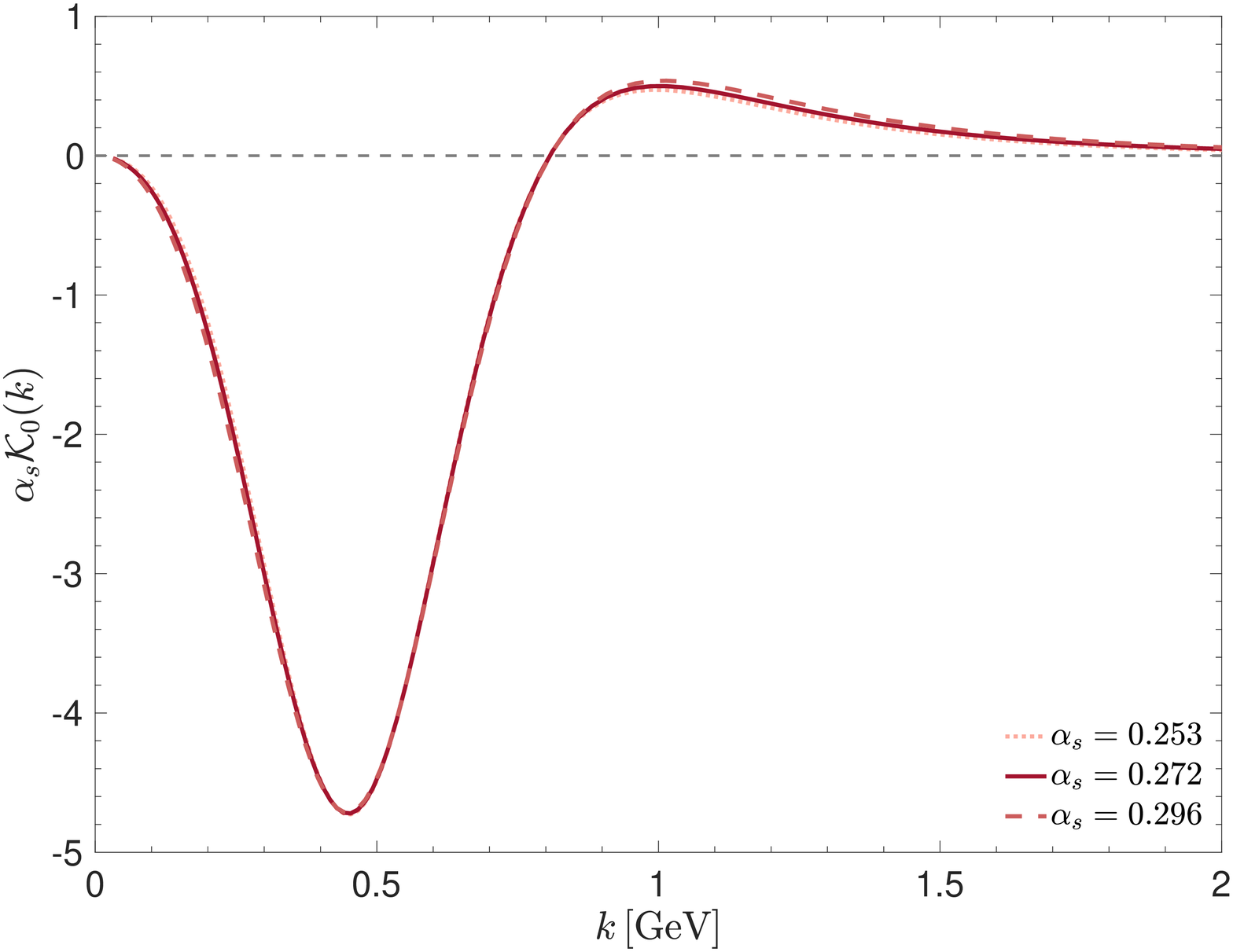}
\end{minipage}
\caption{Left panel:
The values of $\alpha_s$ as a function of the peak height of the ${\cal C}_4(k)$ shown in Fig.~\ref{fig:C34}. The values of $d_1$ are in units of $\mbox{GeV}^2$. 
Right panel: The response of $\alpha_s{\mathcal K}_0(k)$, defined in Eq.~\eqref{K0}, to the
combined variations of ${\cal C}_4(k)$ and $\alpha_s$.} 
\label{fig:C4alpha}
\end{figure}

{\it (iv)} We next analyze the stability of our solutions under variations in the shape  
of ${\mathcal C}_4(k)$. To that end, we solve Eq.~\eqref{eumass} using {\it seven} curves for
${\mathcal C}_4(k)$, which are all located in the shaded band shown in the right panel of Fig.~\ref{fig:C34}. The curves
are obtained by varying  in Eq.~\eqref{c4} the parameter $d_1$,
which controls the height of the peak.  In the left panel of Fig.~\ref{fig:C4alpha} we show the relation between the  maximum value of ${\cal C}_4(k)$ and $\alpha_s$, as $d_1$ is varied  within  the range \mbox{$(4.0-10.0)\,\mbox{GeV}^2$}. It is clear that,
as one reduces the peak range of ${\mathcal C}_4(k)$, the value of $\alpha_s$ increases.  In addition, observe that as the peak of ${\mathcal C}_4(k)$ is approaching the unity (tree-level value),  $\alpha_s$ tends to values higher than $0.3$. Therefore, from this analysis, it is clear that  changes in the peak height (area) of ${\mathcal C}_4(k)$  can be counterbalanced with adjustments in the value of $\alpha_s$,
producing essentially the same solution for $m^2(q)$.

There is a simple way to verify that the same $m^2(q)$ is indeed obtained, by comparing the overall shape of the integrand $\alpha_s{\mathcal K}_0(k)$, defined in Eq.~\eqref{K0}, for different ${\mathcal C}_4(k)$.
In the right panel of Fig.~\ref{fig:C4alpha} we plot $\alpha_s{\mathcal K}_0(k)$, for the variations of ${\mathcal C}_4(k)$ shown in the right panel of Fig.~\ref{fig:C34}. Specifically, the curves are obtained by fixing the values of the pair $(d_1;\alpha_s)$ at  {\it(i)}  $(4.0\,\mbox{GeV}^2;0.296)$ (dashed),  {\it(ii)} $(7.0\,\mbox{GeV}^2;0.272)$ (continuous), and {\it(iii)} $(10.0\,\mbox{GeV}^2;0.253)$ (dotted). It is important to emphasize that
the $\alpha_s$ used for each curve is different, being 
determined from the procedure of solving \1eq{eumass} for each ${\mathcal C}_4(k)$.
As can be clearly seen in Fig.~\ref{fig:C4alpha}, all curves merge into one another; 
plainly, the sets $(\alpha_s, {\mathcal C}_4)$ conspire to eventually 
create the exact same result for $\alpha_s{\mathcal K}_0(k)$. Evidently, since this latter quantity 
remains practically unchanged, the constraint of \1eq{mass0} produces always the same value, \mbox{$m^2(0)=0.14$}\,.

\subsection{\label{sec:ft} Tuning the value of $\alpha_s$}

At first sight, Eq.~\eqref{eumass} appears to be particularly sensitive to changes in $\alpha_s$. As can be observed in the bottom panel of Fig.~\ref{fig:solution}, this sensitivity forces us to tune $\alpha_s$ with three-decimal accuracy in order
to reproduce the lattice value $\Delta(0)$~\cite{Bogolubsky:2007ud}; we remind the reader that the
renormalization (subtraction) point is chosen at $\mu= 4.3$ GeV.  
  
To analyze in some depth the response of Eq.~\eqref{eumass} to  
variations of $\alpha_s$, we next determine the amount by which one may vary it and still
obtain a $\Delta(0)$ lying within the  error bars of the lattice data~\cite{Bogolubsky:2007ud}.

\begin{figure}[t]
\begin{minipage}[b]{0.45\linewidth}
\centering 
\hspace{-1.5cm}
\includegraphics[scale=0.27]{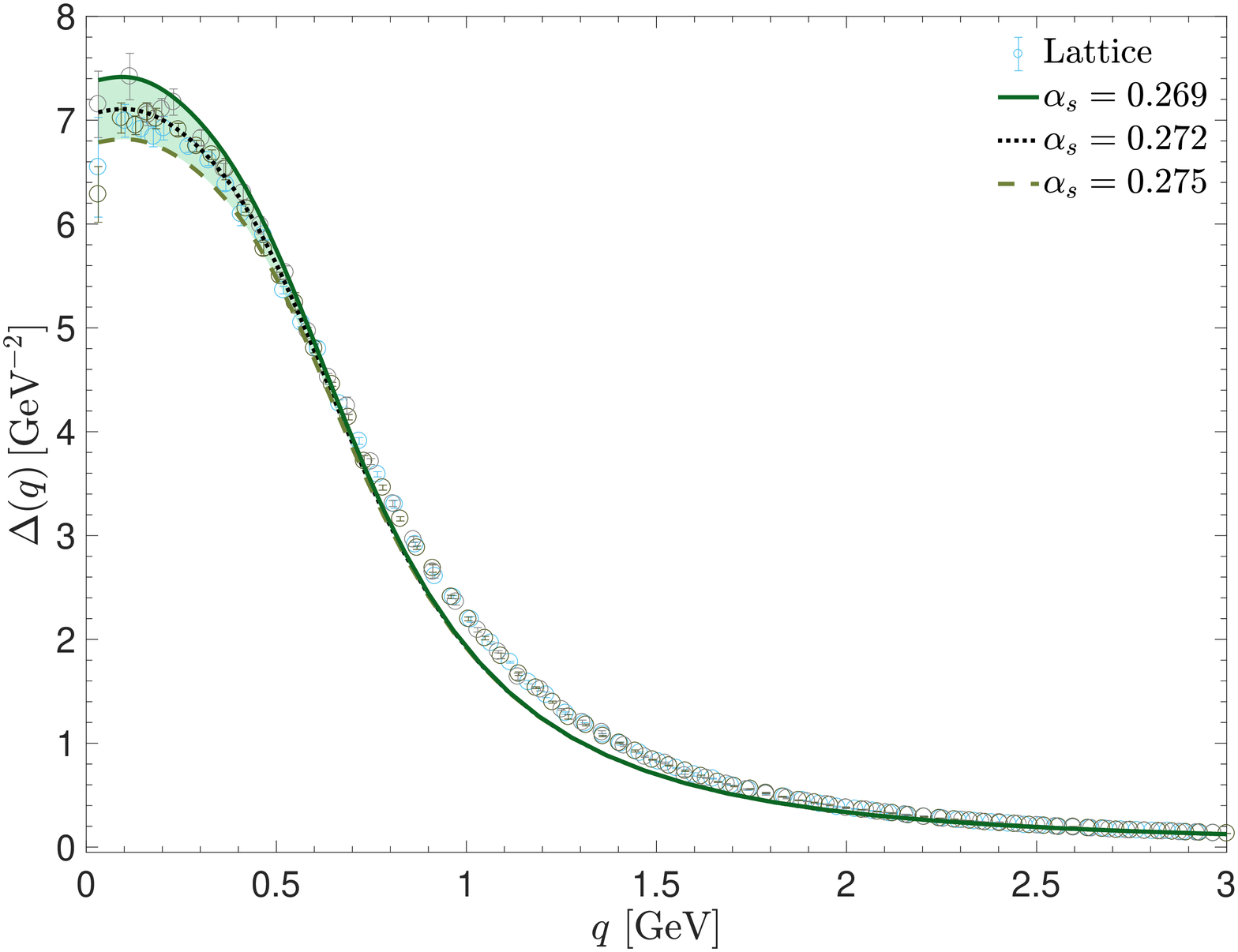}
\end{minipage}
\hspace{0.15cm}
\begin{minipage}[b]{0.45\linewidth}
\includegraphics[scale=0.27]{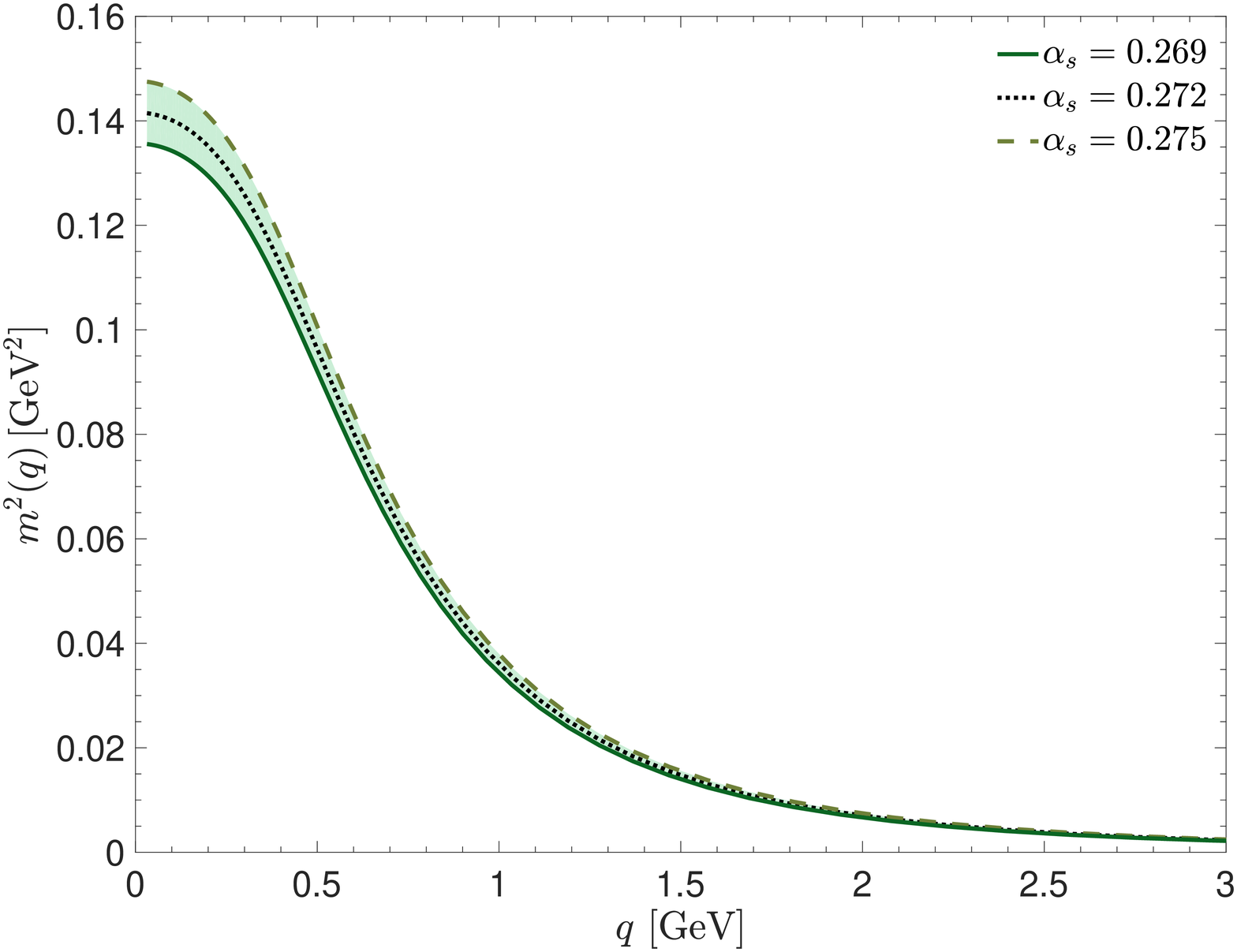}
\end{minipage}
\caption{Left panel: The spreads in the solutions for $\Delta(q)$ when $\alpha_s$ varies $\pm 1\%$. The lattice data are from~\cite{Bogolubsky:2007ud}. Right panel: The corresponding $m^2(q)$.}
\label{fig:spread}
\end{figure}

To that end, we select our result obtained with $J_1(q)$ (the blue dashed dotted
curves in Fig.~\ref{fig:solution}), and vary $\alpha_s$ 
around its central value  $\alpha_s=0.272$. The result of this procedure is
shown in the left panel of Fig.~\ref{fig:spread}, where it can be clearly seen 
that it is possible to cover the spread of the lattice data (in the infrared region) by varying $\alpha_s$ only by $\pm 1\%$.
The corresponding range of solutions for $m^2(q)$ is represented in the right panel of the same figure. 

It turns out that 
the precision in the value of $\alpha_s$ found above may be understood by means of a relatively simple argument. 

In particular, from \1eq{theRG},
\be
{\widehat d}(q) := {\cal R}^2_{\s G}/4\pi = \frac{\alpha_s \Delta(q)}{[1+G(q)]^{2}} \,,
\label{hatd}
\ee
and therefore 
\be
{\widehat d}^{-1}(0) =  \alpha_s^{-1} [1+G(0)]^{2} \Delta^{-1}(0)\,.
\label{hatd0}
\ee

Given that ${\widehat d}^{-1}(0)$ is RGI and has dimensions of mass-squared,
to lowest order it may be written in the form
\be
{\widehat d}^{-1}(0) =  c \mu^2 \exp{(-1/{\tilde b} \alpha_s)} \,,
\label{basrgi}
\ee
with $\tilde b := 4\pi b$, where $b$ is the first coefficient of the Yang-Mills  $\beta$ function, 
$\mu (dg/d\mu) = - b g^3$ (for SU(3), $b=11/16\pi^2$, $\tilde b = 11/4\pi$),  
and  $c$ is a (positive) numerical constant. 

Then, substituting the r.h.s. of (\ref{hatd0}) into the l.h.s. of \1eq{basrgi} yields 
\be
\Delta^{-1}(0) = c \mu^2 [1+G(0)]^{-2} \alpha_s  \exp{(-1/{\tilde b} \alpha_s)} \,.
\label{h1}
\ee
Next, denote by $\delta f$ the variation in the value of a quantity $f$.
If the only source for the variation $\delta\Delta^{-1}(0)$
is the corresponding variation $\delta\alpha_s$ in the value of $\alpha_s$, then 
from \1eq{h1} we obtain
\be
\frac{\delta\alpha_s}{\alpha_s} =  -\sigma \, \frac{\delta \Delta (0)}{\Delta(0)} \,,
\label{h4}
\ee
where we used that $\delta \Delta^{-1}(0)/\Delta^{-1}(0) = - \delta \Delta (0)/\Delta (0)$, and introduced 
\be
\sigma := \frac{{\tilde b} \alpha_s}{1+ {\tilde b} \alpha_s}\,.
\label{h5}
\ee
Note that the minus sign accounts precisely for the tendency shown in the left panel of Fig.~\ref{fig:spread}; namely, an increase (decrease) in the value of  
$\alpha_s$ results in a corresponding decrease (increase) to the value of $\Delta(0)$.  

Taking absolute values, and employing the short-hand notation
$E_f :=\delta f/f$, we have that
\be
E_{\alpha_s} /E_{\Delta(0)} = \sigma\,.
\label{h6}
\ee
From the numerical analysis (see also left panel of Fig.~\ref{fig:spread}), we have that
$E_{\alpha_s} \approx 10^{-2}$, while $E_{\Delta(0)} \approx 4.3 \times 10^{-2}$, so that
$E_{\alpha_s} /E_{\Delta(0)} \approx 0.24$ . On the other hand, 
when we plug into \1eq{h5} the ``central'' value $\alpha_s \approx 0.272$ (at $\mu= 4.3$ GeV)  
we find that $\sigma \approx 0.19$, concluding that \1eq{h6} is satisfied reasonably well.
This simple ballpark estimate seems to indicate 
that the required tuning in the value of $\alpha_s$ is compatible with what one would expect
on general grounds, and is, in that sense, fairly natural.

\section{\label{sec:conc} Discussion and Conclusions}

In this work we have demonstrated how the nonlinear treatment of the
gluon gap equation, in conjunction with an effective implementation of 
multiplicative renormalization, fixes the value of the emergent gluonic scale,
and gives rise to positive-definite and monotonically decreasing
running gluon masses. 
In particular, the analysis presented relies on the following pivotal points:

({\it i}) The nonlinearization of the equation proceeds by implementing \1eq{eq:gluon_m_J}
for the gluon propagators appearing in it; this substitution, in turn, 
introduces the unknown function $m^2(q)$
in the corresponding denominators, thus eliminating the freedom of rescaling the solutions.

({\it ii}) For the kinetic term $J(q)$,  entering into the mass equation after the use of 
\1eq{eq:gluon_m_J}, we employ physically motivated {\it Ans\"atze} which capture its salient features,
and are further refined during the iterative numerical procedure. 

({\it iii}) An effective approach to multiplicative renormalization,
inspired from analogous studies in the quark sector of the theory,
has been implemented, which 
introduces into the mass equation two additional form-factors, one for the 
three-gluon and one for the four-gluon vertex.

({\it iv})
Due to the inclusion of these 
form factors, the ``competition''
between the one- and two-loop terms comprising the mass equation
(carrying a relative minus sign) is tilted slightly in favor of the latter.
In particular, the infrared suppression of the three-gluon vertex
reduces the size of the one-loop term, while the enhancement of the four-gluon form factor boosts
up the two-loop contribution, 
such that, eventually, solutions with the desired properties are obtained.

({\it v}) In various demonstrations throughout this article, and especially in 
Sec.~\ref{sec:multren}, we have relied extensively on special RGI combinations, 
whose use renders the relevant constructions considerably more transparent.

It is interesting to comment on the relevance of the quantity $-{\dot m}^2(q)$,
plotted in the left panel of Fig.~\ref{fig:massB}.
As has been explained in a series of works (see \eg ~\cite{Aguilar:2011xe,Binosi:2017rwj}),
on theoretical grounds this quantity is exactly equal to the Bethe-Salpeter amplitude
that controls the formation of the massless excitations that trigger the
Schwinger mechanism, and the subsequent generation of a gluon mass. Evidently, the levels of accuracy achieved  
in fulfilling this equality provide a highly nontrivial check
of the entire mechanism, in general, and of the veracity of the approximations employed, in particular. 
A direct comparison between Fig.~\ref{fig:massB} of the present work and Fig.~5 of~\cite{Binosi:2017rwj}
reveals that while the qualitative behavior is similar, the corresponding maxima are relatively
further apart [340 MeV and 1 GeV, respectively]. Note, however, that all existing analyses of this 
particular Bethe-Salpeter equation are also linear, in the sense that, as in the case of the
mass equation, the gluon propagators entering in it were treated as external quantities.
It turns out that a nonlinear approach to this problem amounts
to solving a rather complicated integro-differential equation, whose 
numerical treatment is already underway. 

We emphasize that all ingredients used in the present analysis have been
renormalized at $\mu = 4.3$ GeV; therefore, it is understood that all non-RGI results obtained,
such as the $m^2(q)$ and the value of $\alpha_s$ employed, are valid for this particular choice of $\mu$. 
It would certainly be important to establish the response and overall stability
of the mass equation under changes in the value of $\mu$. 
Even though we will not pursue this issue any further here,
we outline the general method that one should adopt~\cite{Aguilar:2010gm}; the basic steps
may be summarized as follows: ({\it a}) In general, dimensionless quantities, $f(k)$,
such as ${\cal C}_3(k)$ and ${\cal C}_4(k)$, 
whose form is computed (or assumed) at a scale $\mu_1$, are rescaled to a different point  $\mu_2$ 
according to $f(k,\mu_2) = f(k,\mu_1)/f(\mu_2,\mu_1)$. 
On the other hand, the gluon propagator corresponding to the lattice result renormalized at $\mu_2$
is obtained from the corresponding result at $\mu_1$ through $\Delta(k,\mu_2) = \Delta(k,\mu_1)/\mu_2^2\Delta(\mu_2,\mu_1)$,
({\it b}) The curves of  ${\cal C}_3(k,\mu_2)$ and ${\cal C}_4(k,\mu_2)$ are to be 
substituted into the mass equation, and the new value of $\alpha_s = \alpha_s(\mu_2)$ must be determined, such that
the resulting $m^2(0,\mu_2) = \Delta^{-1}(0,\mu_2)$.
({\it c}) The repetition of these steps for a set of $\{\mu_i\}$
will essentially furnish the evolution of $\alpha_s$ that is required by the
gluon mass equation; this curve, in turn, must be compared with the evolution of $\alpha_s$
expected on general grounds, and the level of agreement established. 

As already mentioned, the kinetic term $J(q)$ of the gluon propagator satisfies
its own dynamical equation, which, due to the technical complexities associated with
several of its ingredients,
has not been presented in the literature. However, recent progress 
accomplished in various fronts, and especially our firmer knowledge on
the behaviour of the three-gluon vertex, seems to bring this task well within our reach.
In fact, it would be clearly desirable to eventually solve the {\it coupled system} of equations for
$m^2(q)$ and  $J(q)$, and establish how closely the lattice results
for both the gluon propagator and the 
three-gluon vertex may be reproduced.  
Calculations in that directions are already in progress, 
and we hope to present new results in the near future. 

\acknowledgments 
J.~P. thanks Jan Pawlowski, Craig Roberts, and Jos\'{e} Rodr{\'i}guez-Quintero
for several stimulating discussions.  
The research of J.~P. is supported by the 
The Spanish Ministry of Economy and Competitiveness (MINECO) under grants FPA2017-84543-P and SEV-2014-0398.
The work of  A.~C.~A. and M.~N.~F. are supported by the Brazilian National Council for Scientific and Technological Development (CNPq) under the grants 305815/2015,   142226/2016-5, and 464898/2014-5 (INCT-FNA). A.~C.~A. and C.~T.~F. also acknowledge the financial support
from  S\~{a}o Paulo Research Foundation (FAPESP) through the projects  2017/07595-0, 
2017/05685-2, 2016/11894-0, and 2018/09684-3.  This study was financed in part by the Coordena\c{c}\~{a}o de Aperfei\c{c}oamento de Pessoal de N\'{\i}vel Superior - Brasil (CAPES) - Finance Code 001 (M.~N.~F.).


\end{document}